\documentclass[aps,rmp,reprint,superscriptaddress]{revtex4-2}
\usepackage{CJK}
\usepackage{enumerate,appendix}
\usepackage{commath,braket}
\usepackage{color,calc,graphicx}
\usepackage[usenames,dvipsnames,svgnames,table,cmyk,hyperref]{xcolor}
\usepackage{epstopdf}
\usepackage[charter,cal=cmcal,sfscaled=false]{mathdesign}
\definecolor{oeawblue}{cmyk}{0.9,0.68,0,0}
\definecolor{iqoqiblue}{cmyk}{0.76,0.11,0,0}
\usepackage{hyperref}

\hypersetup{
  pdftitle={Optimized detection of high-dimensional entanglement},
  pdfstartview=Fit,
  pdfpagelayout=SinglePage,
  colorlinks,
  citecolor=oeawblue,
  linkcolor=oeawblue,
  urlcolor=oeawblue}
\usepackage[figure]{hypcap}

\def\tr{\mbox{tr}}
\def\id{{\mathbb I}}
\def\R{\mathbb{R}}
\def\C{{\mathbb C}}

\newcommand{\probi}{p_\texttt{I}}
\newcommand{\probii}{p_\texttt{II}}
\newcommand{\rhot}[1]{\rho_{\text{#1}}}
\newcommand{\sigt}[1]{\sigma_{\text{#1}}}
\def\proj#1{\ket{#1}\!\bra{#1}}

\begin{document}

\title{Optimized Detection of High-Dimensional Entanglement}

\author{Xiao-Min Hu}
\email{These two authors contributed equally to this work.}
\affiliation{CAS Key Laboratory of Quantum Information, University of Science and Technology of China, Hefei 230026, China}
\affiliation{CAS Center For Excellence in Quantum Information and Quantum Physics, University of Science and Technology of China, Hefei, 230026, People's Republic of China}
\author{Wen-Bo Xing}
\email{These two authors contributed equally to this work.}
\affiliation{CAS Key Laboratory of Quantum Information, University of Science and Technology of China, Hefei 230026, China}
\affiliation{CAS Center For Excellence in Quantum Information and Quantum Physics, University of Science and Technology of China, Hefei, 230026, People's Republic of China}
\author{Yu Guo}
\affiliation{CAS Key Laboratory of Quantum Information, University of Science and Technology of China, Hefei 230026, China}
\affiliation{CAS Center For Excellence in Quantum Information and Quantum Physics, University of Science and Technology of China, Hefei, 230026, People's Republic of China}
\author{Mirjam Weilenmann}
\affiliation{Institute for Quantum Optics and Quantum Information--IQOQI Vienna, Austrian Academy of Sciences, Boltzmanngasse 3, 1090 Vienna, Austria}
\author{Edgar A. Aguilar}
\affiliation{Institute for Quantum Optics and Quantum Information--IQOQI Vienna, Austrian Academy of Sciences, Boltzmanngasse 3, 1090 Vienna, Austria}
\author{Xiaoqin Gao}
\email{xgao5@uottawa.ca}
\affiliation{Institute for Quantum Optics and Quantum Information--IQOQI Vienna, Austrian Academy of Sciences, Boltzmanngasse 3, 1090 Vienna, Austria}
\affiliation{Vienna Center for Quantum Science and Technology (VCQ), Faculty of Physics, University of Vienna, Boltzmanngasse 5, 1090 Vienna, Austria}
\affiliation{Department of Physics, University of Ottawa, Advanced Research Complex, 25 Templeton Street, K1N 6N5, Ottawa, ON, Canada}
\author{Bi-Heng Liu}
\email{bhliu@ustc.edu.cn}
\affiliation{CAS Key Laboratory of Quantum Information, University of Science and Technology of China, Hefei 230026, China}
\affiliation{CAS Center For Excellence in Quantum Information and Quantum Physics, University of Science and Technology of China, Hefei, 230026, People's Republic of China}
\author{Yun-Feng Huang}
\affiliation{CAS Key Laboratory of Quantum Information, University of Science and Technology of China, Hefei 230026, China}
\affiliation{CAS Center For Excellence in Quantum Information and Quantum Physics, University of Science and Technology of China, Hefei, 230026, People's Republic of China}
\author{Chuan-Feng Li}
\email{cfli@ustc.edu.cn}
\affiliation{CAS Key Laboratory of Quantum Information, University of Science and Technology of China, Hefei 230026, China}
\affiliation{CAS Center For Excellence in Quantum Information and Quantum Physics, University of Science and Technology of China, Hefei, 230026, People's Republic of China}
\author{Guang-Can Guo}
\affiliation{CAS Key Laboratory of Quantum Information, University of Science and Technology of China, Hefei 230026, China}
\affiliation{CAS Center For Excellence in Quantum Information and Quantum Physics, University of Science and Technology of China, Hefei, 230026, People's Republic of China}
\author{Zizhu Wang}
\email{zizhu@uestc.edu.cn}
\affiliation{Institute of Fundamental and Frontier Sciences, University of Electronic Science and Technology of China, Chengdu 610054, China}
\author{Miguel Navascu\'es}
\email{miguel.navascues@oeaw.ac.at}
\affiliation{Institute for Quantum Optics and Quantum Information--IQOQI Vienna, Austrian Academy of Sciences, Boltzmanngasse 3, 1090 Vienna, Austria}

\begin{abstract}
Entanglement detection is one of the most conventional tasks in quantum information processing. While most experimental demonstrations of high-dimensional entanglement rely on fidelity-based witnesses, these are powerless to detect entanglement within a large class of entangled quantum states, the so-called \emph{unfaithful} states. In this paper, we introduce a highly flexible automated method to construct optimal tests for entanglement detection given a bipartite target state of arbitrary dimension, faithful or unfaithful, and a set of local measurement operators. By restricting the number or complexity of the considered measurement settings, our method outputs the most convenient protocol which can be implemented using a wide range of experimental techniques such as photons, superconducting qudits, cold atoms or trapped ions. With an experimental quantum optics setup that can prepare and measure arbitrary high-dimensional mixed states, we implement some $3$-setting protocols generated by our method. These protocols allow us to experimentally certify $2$- and $3$-unfaithful entanglement in $4$-dimensional photonic states, some of which contain well above 50\% of noise.
\end{abstract}

\maketitle

Entanglement is the bedrock of most quantum information processing protocols~\cite{Horodecki2009RMP,guhne20091}. It is a key resource in quantum teleportation \cite{teleportation}, entanglement-based quantum key distribution (QKD) \cite{QKD} and quantum communication complexity~\cite{Cleve2010RMP}. Generating high-quality entangled states and detecting them reliably is a crucial prerequisite to conduct any quantum communication task. However, entanglement detection is a computationally hard problem for high-dimensional systems~\cite{strong_NP_sep,NP_sep}. There are only a few known general methods, all of which come with high computational costs~\cite{DPS1,DPS2,DPS3,Toth2015}. Most experimental entanglement detection protocols, especially those which aim to detect high-dimensional entanglement, use linear witnesses based on the fidelity between the generated state and a (pure) target state, see, e.g.,~\cite{Bavaresco2018NP,Hu2020PRL,zhong201812,malik2016multi}. Fidelity-based witnesses have been shown to work well for the states generated in current experiments, with only a constant number of measurement settings needed to measure each witness~\cite{FlammiaLiu2011PRL}. However, the recent discovery of \emph{unfaithful} entanglement~\cite{Mirjam2020}, which cannot be detected by fidelity-based witnesses, has changed the status quo.

Unfaithful entanglement is not the exception, but the norm: almost all high-dimensional bipartite entangled states are unfaithful~\cite{Mirjam2020}. This situation posed a conundrum for theorists and experimentalists alike: how to supplement fidelity-based witnesses with experiment-friendly protocols which are capable of detecting unfaithful entanglement? Here we tackle this problem by designing a method to automatically search for optimal protocols for certifying high-dimensional bipartite entanglement, including the unfaithful kind, with only a few local measurements. Specifically, using any bipartite target state $\rhot{AB}$, a set of local measurement operators $\{N^x\}_x$, and the entanglement dimension $D+1$ of $\rhot{AB}$ to be certified as input, our method constructs a two-element positive operator-valued measure (POVM), $M=(M_{\texttt{C}}, M_{\texttt{U}})$, where outcome \texttt{C} stands for certified $D+1$-dimensional entanglement and \texttt{U} for `uncertified'. This POVM can be conducted through a one-way local operations and classical communication (LOCC) protocol, and the probability of mistakenly reporting outcome \texttt{C} when a state of Schmidt rank $D$ or lower was prepared or outcome \texttt{U} when the target state $\rhot{AB}$ was prepared is minimal as guaranteed by convex optimization theory. If we conduct several experimental implementations of the protocol, then the number of occurrences of outcome \texttt{C} can be used to certify, with high statistical confidence, that the generated state $\rhot{exp}$ has Schmidt rank at least $D+1$, even if the protocol's measurement settings are far from tomographically complete. These protocols can be implemented in a wide variety of physical systems, especially where the presence of noise in high-dimensional entangled states renders them unfaithful thus making their certification with fidelity-based witnesses impossible.

We experimentally tested our method with a dozen 4-dimensional bipartite target states, each belonging to one of two groups. The archetype of the first group has 3-dimensional entanglement (3-entangled) but is 3-unfaithful and the second group is 2-entangled but 2-unfaithful. Using our method, the entanglement dimension of the first group can be certified with 3 commonly used measurement settings per side and estimating 22 to 44 different probabilities. For the 2-unfaithful states, a different set of 3 measurement settings per side and 89 probabilities are needed. 

\textit{Method for generating optimal protocols for high-dimensional entanglement detection}---
We start from the following premise which covers the most basic experimental scenario: two parties, Alice and Bob, want to certify the entanglement dimension of a shared quantum state $\rhot{AB}$ with local dimension $d$. Each of them can perform $m$ local measurements in their part of the lab, with each measurement producing one of $d$ possible outcomes. We can hence identify each measurement $x\in\{1,...,m\}$ by the set of POVM elements $N^x\equiv \{N_{a|x}: a\in \{1,...,d\}\}$. We further allow Alice and Bob to carry out 1-way LOCC protocols. Namely, we allow Alice to measure first and then communicate her measurement setting $x$ and outcome $a$ to Bob. With this information, Bob decides which measurement $y$ to conduct in his lab, call $b$ his measurement result. Alice and Bob's guess on the entanglement of their shared state will be a non-deterministic function of $a,b,x,y$.

More formally, call $S_D$ the set of all quantum states $\sigma_{AB}$ that admit a decomposition of the form:
\begin{equation}
\sigt{AB}=\sum_i \lambda_i\proj{\psi_i}_{AB},
\end{equation}
\noindent where each state $\ket{\psi_i}$ has Schmidt rank $D$ or lower, and the weights $\{\lambda_i\}$ satisfy $\lambda_i\geq 0$, for all $i$ and $\sum_i \lambda_i=1$. Any state that does not belong to $S_D$ has entanglement dimension at least $D+1$.

Let $\rhot{AB}$ be such a state: $\rhot{AB}\not\in S_D$, and let $\probi, \probii\in [0,1]$. We first consider $1$-shot $1$-way LOCC measurement protocols for Alice and Bob, with possible outcomes $\texttt{C}$ (certified) and $\texttt{U}$ (uncertified), such that

\begin{enumerate}
\item
If the state $\sigma$ shared by Alice and Bob belongs to $S_D$, the probability that they output $\texttt{C}$ is bounded by $\probi$.
\item
If the state shared by Alice and Bob is indeed the target state $\rhot{AB}$, the probability that Alice and Bob output $\texttt{U}$ is bounded by $\probii$.

\end{enumerate}

The conditions above constitute a hypothesis test, with $\probi$ and $\probii$ playing the roles of type-I (false positive) and type-II (false negative) errors. Using the newly developed quantum preparation games~\cite{WeilenmannPrepGames}, we can recast our search for $1$-shot $1$-way LOCC protocols that minimise the sum $\probi+\probii$ into the following optimization problem (see Appendix~\ref{app:theory} for more theoretical analysis):
\begin{align}
\text{Minimize} \quad &\probi+\probii\nonumber\\
\text{subject to} \quad &\tr(M_{\texttt{U}}\rhot{AB})=\probii,\nonumber\\
&\Pi_D\left(\probi\id_{AB}-M_{\texttt{C}}\right)\Pi_D^\dagger=\nonumber\\
&\Lambda^1_{AA'B'B}+(\Lambda^2_{AA'B'B})^{T_{BB'}}+\lambda\left(\Pi_D\Pi_D^\dagger-\frac{\id}{D}\right),\nonumber\\
&M_{\texttt{C}}=\sum_{x,y,a,b}P\left(x,y,c=\texttt{C}|a,b\right)N_{a|x}\otimes N_{b|y},\nonumber\\
&M_{\texttt{U}}=\id_{AB}-M_{\texttt{C}},\nonumber\\
&\Lambda^1,\Lambda^2\succeq 0,\; P(x,y, c|a,b)\geq 0, \nonumber\\
&\sum_c P(x,y, c|a,b)=P(x,y|a),\nonumber\\
&\sum_yP(x,y|a)=P(x),\sum_x P(x)=1.
\label{main_SDP}
\end{align}
\noindent Here auxiliary systems $A', B'$ are assumed to have dimension $D$, and $\Pi_D =\id_{A}\otimes\ket{\psi_D^+}_{A'B'}\otimes\id_B$, with $\ket{\psi_D^+}=\sum_{i=1}^D\ket{i}\ket{i}$ being the non-normalised maximally entangled state in $\C^D\times\C^D$. $T_{BB'}$ denotes the partial transpose over systems $BB'$. The optimization variables $P(x,y,c|a,b)$ represent a collection of probability distributions, with $x,y\in\{1,...,m\}$, $a,b\in\{1,...,d\}$, $c\in\{\texttt{C},\texttt{U}\}$. The meaning of $\Lambda^1,\Lambda^2$ and a more complete derivation can be found in Appendix~\ref{app:theory}.

\begin{figure*}[htbp!]
\centering
\includegraphics[width=0.8\textwidth]{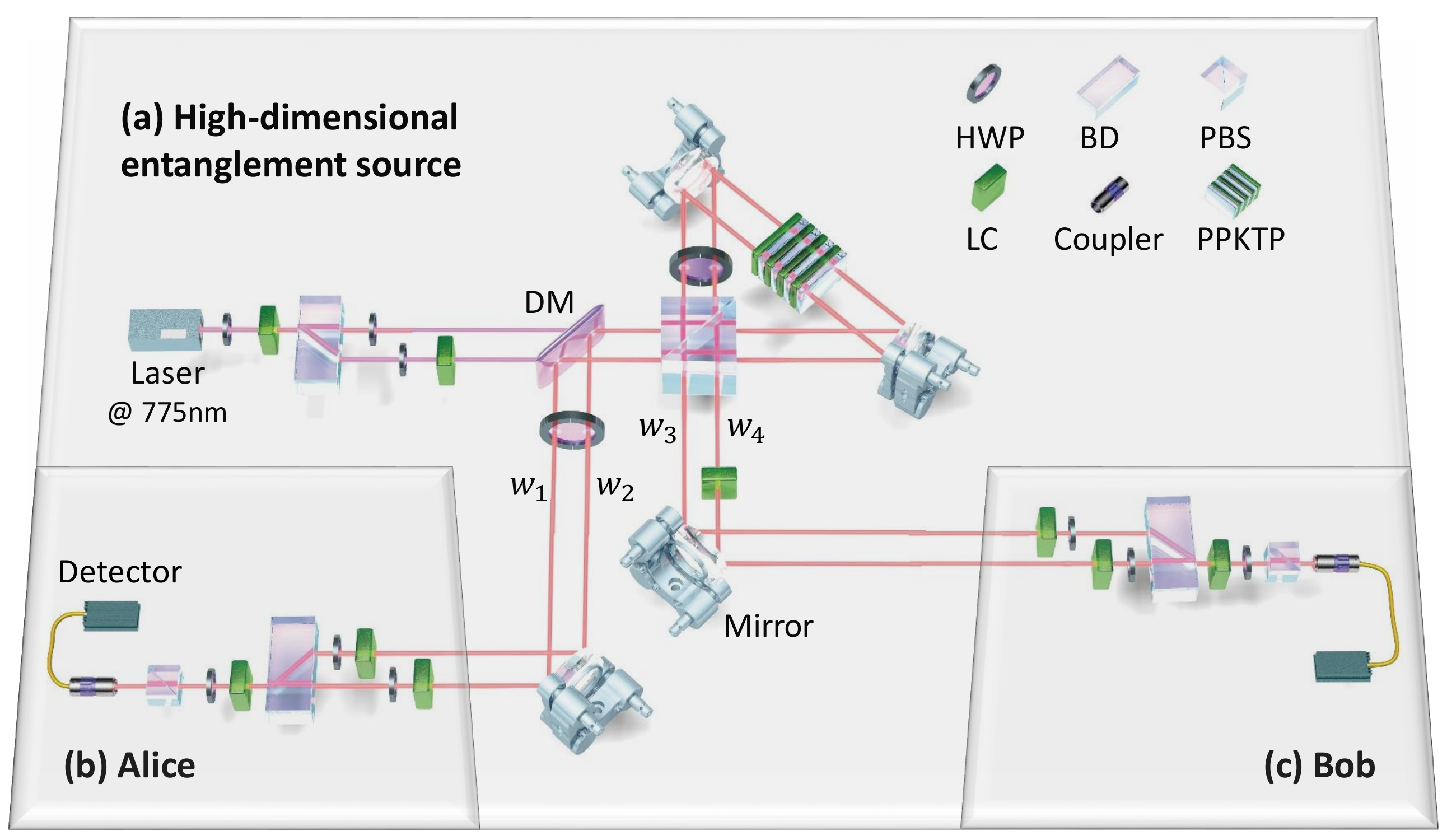}
\caption{Experimental setup for generating the unfaithful states. (a) Preparation of a four-dimensional entanglement source @ $1550~nm$. After two lenses (not shown), the continuous-wave (CW) laser beam (@ $775~nm$ with power @ $200~mW$) is focused into a small beam with waist radius of $\approx0.7~mm$. With a beam-displacer (BD), the beam is separated into two paths. Then, these two beams are injected into a Sagnac interferometer, which generates a two-photon polarization-based spontaneous parametric down-conversion(SPDC) entangled state $\frac{1}{\sqrt{2}}(|HV\rangle+|VH\rangle)$ in each path~\cite{Hu2018} by a type-II PPKTP crystal ($1 mm\times7 mm\times10 mm$, the polling period is $46.2~\mu m$, and the temperature is set at $35^{\circ}C$. In our experiment, the full width at half maximum (FWHM) of the down-converted photons is $\approx2~nm$.). After a half-wave plate (HWP) @$45^{\circ}$, each entangled state is rotated to $\frac{1}{\sqrt{2}}(|HH\rangle+|VV\rangle)$, which is encoded into the paths $w_1 (w_3)$ and $w_2 (w_4)$, respectively.
If we define the path $w_1 (w_3)$ with $H$-polarized photon as $\ket{0}$, $w_1 (w_3)$ with $V$-polarized photon as $\ket{1}$, $w_2 (w_4)$ with $H$-polarized photon as $\ket{2}$, and $w_2 (w_4)$ with $V$-polarized photon as $\ket{3}$, then a four-dimensional maximally entangled state is generated.
Afterwards, to prepare the state $\rhot{UNF}^{p}$, we need to change the angles of first three HWPs and liquid crystals (LCs) together with their voltages. If we only consider the paths $w_1$ and $w_3$, then we obtain the state $\ket{\Psi_{2}}$.
An additional light source is introduced as white noise mixed with the pure state $\ket{\Psi_{2}}$ to generate the states $\rhot{ISO2}^p$. (b) and (c), setups for Alice's and Bob's local measurements. Each party has three LCs, three HWPs, a BD, a PBS and a single-photon superconducting detector. By adjusting the voltages of the LCs and the angles of the HWPs, different measurement bases can be implemented.}
\label{fig:exp_setup}
\end{figure*}

In effect, Eq.~(\ref{main_SDP}) describes the following problem: given the state $\rhot{AB}$ shared by Alice and Bob, together with their allowed local measurements $N_{a|x},\, N_{b|y}$, what is the best $1$-way LOCC strategy that maximises the chance they can correctly certify the entanglement dimension of $\rhot{AB}$ to be $D+1$ while at the same time minimises the chance they make mistakes? The solution to this problem is a feasible sum of $\probi$ and $\probii$, with the operational description of the protocol achieving this sum encoded in the minimizer $P^\star(x,y,c|a,b)$. The optimization algorithm also returns the optimal $\probi$ satisfying the first condition of the hypothesis test. This allows us to compute the $p$-value of the null hypothesis after $n$ experimental repetitions of the protocol. To solve Eq.~(\ref{main_SDP}), which is a semidefinite program (SDP), any number of readily-available solvers such as MOSEK~\cite{mosek} can be used, which not only output the numerical solution of the problem, but also rigorous upper and lower bounds on its optimal value. The only practical limitation is the physical dimension $d$ and the entanglement dimension $D$, even though the method itself is valid for any $d$ and $D$. The protocols used in our experiment were computed on a normal desktop computer in less than 30 seconds. In fact, problem~(\ref{main_SDP}) is just the first level in a hierarchy of SDPs that output feasible 1-way LOCC protocols with provable type-I error $\probi$ and decreasing sum $\probi+\probii$. In the infinite limit, the hierarchy returns the measurement protocol that minimizes $\probi+\probii$. For small dimensions, the first level already provides good enough protocols for $D+1$-dimensional entanglement detection. Additional information such as the complete description of the hierarchy, the meaning of $\Lambda^1,\Lambda^2$ in~(\ref{main_SDP}), and a benchmark of the method with random target states and different sets of measurement settings can be found in Appendix~\ref{app:theory}.


Since our experimental setup does not allow us to switch the measurement settings dynamically, the LOCC protocols are not implemented directly. Instead, under the assumption that the source produces the same target state $\rho^{\text{exp}}$ in every experimental round, we estimate the operator averages $\langle N_{a|x}\otimes N_{b|y}\rangle_{\text{exp}}$, for $a,b,x,y$. With these averages we compute the probability $\probii^{\text{exp}}$ of obtaining result $\texttt{U}$, had we implemented the optimal 1-way LOCC protocol $P^\star(x,y,c|a,b)$ once. If $\probii^{\text{exp}}$ is such that $\probi+\probii^{\text{exp}}<1$, then we can conclude that $\rho^{\text{exp}}$ has entanglement dimension at least $D+1$. 

\begin{figure*}[htbp!]
\begin{center}
\includegraphics [width= 1.8\columnwidth]{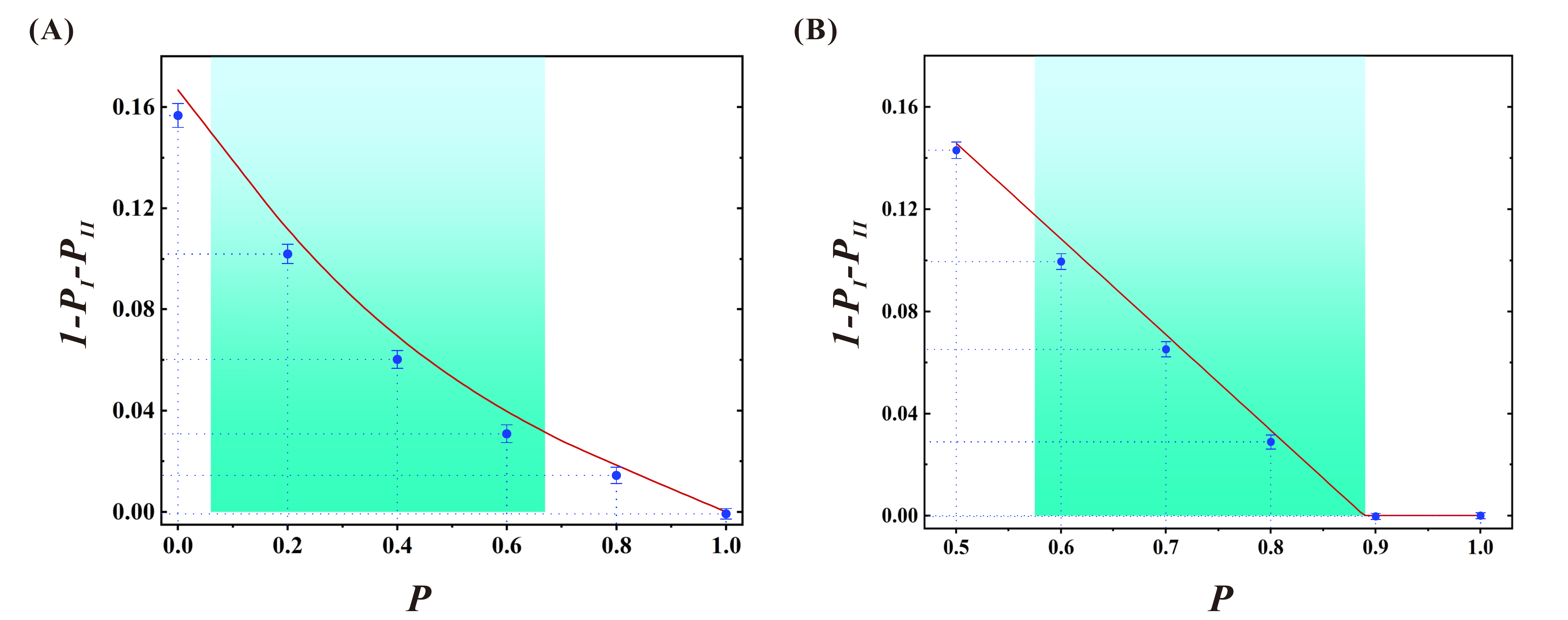}
\end{center}
\caption{Experimental results of the optimal $1-\probi-\probii$ for states. 
Experimental values for $p=0,\,0.2,\,0.4,\,0.6,\,0.8$, and $1$ of state $\rhot{UNF}^p$ in (a) and $p=0.5,\,0.6,\,0.7,\,0.8,\,0.9$, and $1$ of state $\rhot{ISO}^p$ in (b) are shown in blue.
The solid red lines represent optimal values computed by the SDP solver.
States in the green regions are 3-unfaithful in (a) and 2-unfaithful in (b). 
The entanglement dimension is certified when $1-\probi-\probii>0$.}
\label{fig:unf3}
\end{figure*}


\textit{Experimental certification of unfaithful entanglement}---
The 2 types of target states we selected for experimental testing are:
\begin{align}
\rhot{UNF}^p&=(1-p)\proj{\Psi_{3}}+p\left(\frac{\ket{23}+\ket{32}}{\sqrt{2}}\right)\left(\frac{\bra{23}+\bra{32}}{\sqrt{2}}\right)\\
\rhot{ISO2}^p&=(1-p)\proj{\Psi_{2}}+\frac{p}{16}\mathbb{I}_{4} \otimes \mathbb{I}_4,
\end{align}
with $\ket{\Psi_{d}}$ denoting the $d$-dimensional maximally entangled state $\ket{\Psi_{d}}=(1/\sqrt{d})\sum_{i=0}^{d-1}\ket{ii}$. The state $\rhot{UNF}^p$ is a mixture of pure states. Experimental methods which allow preparing this kind of high fidelity mixed states would be particularly useful because currently there are very few feasible options. We here present a general method that mixes arbitrary bipartite pure states by rapidly switching the electro-optical modulation elements (See Appendix~\ref{app:exp} for more information).

Using the computational tests SDP1 and SDP2 in~\cite{Mirjam2020}, we can certify that $\rhot{UNF}^{0.2}$,$\rhot{UNF}^{0.4}$,$\rhot{UNF}^{0.6}$ is 3-entangled but 3-unfaithful while the states $\rhot{ISO2}^{0.6},\,\rhot{ISO2}^{0.7},\,\rhot{ISO2}^{0.8}$ are 2-entangled but 2-unfaithful. The states $\rhot{ISO2}^p$ model the situation where high-dimensional noise creeps into the preparation of entangled qubits, making them unfaithful thus unsuitable for fidelity-based witnesses. For such states, our method shows that entanglement detection is possible with as much as over $80\%$ of noise, which is confirmed by our experiment as shown in Fig.~\ref{fig:unf3}.


The target states are generated by the setup depicted in Fig.~\ref{fig:exp_setup}. For the state $\rhot{UNF}^{p}$, we first generate the two pure states $\ket{\Psi_{3}}$ and $\frac{1}{\sqrt{2}}(\ket{23}+\ket{32})$ using polarization-based path degree of freedom of photons, as shown in Fig.~\ref{fig:exp_setup}(a), then mix them by rapidly switching the voltages of liquid crystals (LCs) (more information can be found in Appendix~\ref{app:exp}). LCs play an important role in our mixed state preparation. The relative phase between the $H$- and $V$-polarised photons can be changed by adjusting the voltage applied to the LC, which makes it behave like an HWP. The relative phase between $H$ and $V$ varies from 0, when the applied voltage is $V_I$, to $\pi$, for $V_{\pi}$. The polarization is stabilized by keeping $V_{I}$ stable for the three LCs. By controlling the loading time of different phases, we can prepare $\rhot{UNF}^p$ with different values of $p$ (more information can be found in Appendix~\ref{app:exp}).

To generate the states $\rhot{ISO2}^p$, we mix a two-dimensional maximally entangled state $\ket{\Psi_{2}}$ generated by the pump with 4-dimensional white noise emitted by independent sources~\cite{Ecker2019,Hu2020PRL} (more information can be found in Appendix~\ref{app:exp}).

After the states are generated, we performed tomographic reconstructions of their density matrices using maximum-likelihood estimation (MLE)~\cite{thew2002qudit,hu2016experimental}. The reconstructed density matrices are used to certify that the generated states are indeed unfaithful.

\begin{table*}[htbp!]
\centering
\caption{Experimental results. $\mathcal{N_S}$ and $\mathcal{N_P}$ are the number of settings and probabilities, respectively. $D$ is the entanglement dimension.}
\begin{tabular}{ |c|c|c|c|c|c| }
 \hline
 Target state & $1-\probi-\probii$ & $1-\probi-\probii^{\text{exp}}$& $\mathcal{N_S}$ & $\mathcal{N_P}$ & $D$ \\ 
  \hline
 $\rhot{UNF}^{0}$ & $0.167$ & $0.160\pm 0.005$ & 3 & $38$ & $3$\\ [4pt]
  \hline
 $\rhot{UNF}^{0.2}$ & $0.112$ & $0.102\pm 0.004$ & 3 & $22$ & $3$\\ [4pt]
  \hline
 $\rhot{UNF}^{0.4}$ & $0.070$ & $0.069\pm 0.004$ & 3 & $26$ & $3$\\ [4pt]
  \hline
 $\rhot{UNF}^{0.6}$ & $0.040$ & $0.031\pm 0.003$ & 3 &$26$ & $3$\\ [4pt]
  \hline
 $\rhot{UNF}^{0.8}$ & $0.018$ & $0.014\pm 0.003$ & 3 &$44$ & $3$\\ [4pt]
   \hline
 $\rhot{UNF}^{1.0}$ & $0$ & $-0.000\pm 0.002$ & 3 &$144$ & N/A \\ [5pt]
 \hline
 $\rhot{ISO2}^{0.5}$ & $0.146$ & $0.143\pm 0.003$ & 3 &$89$ & $2$\\ [4pt]
 \hline
 $\rhot{ISO2}^{0.6}$ & $0.108$ & $0.100\pm 0.003$ & 3 &$89$ & $2$\\ [4pt]
 \hline
 $\rhot{ISO2}^{0.7}$ & $0.071$ & $0.065\pm 0.003$ & 3 &$89$ & $2$\\ [4pt]
 \hline
 $\rhot{ISO2}^{0.8}$ & $0.033$ & $0.029\pm 0.003$ & 3 &$89$ & $2$\\ [4pt]
 \hline
 $\rhot{ISO2}^{0.9}$ & $0$ & $0.001\pm 0.001$ & 3 &$144$ & N/A \\ [4pt]
 \hline
 $\rhot{ISO2}^{1.0}$ & $0$ & $-0.000\pm 0.001$ & 3 &$144$ & N/A \\ [4pt]
 \hline
 $\Phi_{4}$ & $0.084$ & $0.073\pm 0.005$ & 2 &$8$ & 4\\ [4pt]
 \hline
\end{tabular}

\label{table_results}
\end{table*}

Only 3 measurement settings are needed in our experiment to certify the entanglement dimensions of $\rhot{UNF}^p$ and $\rhot{ISO2}^p$. The optimal $1$-shot $1$-way LOCC protocol using these measurement settings found by the SDP solver, in the form of a set of probability distributions $P(x,y,c|a,b)$, allow us to compute $\probii^{\text{exp}}$ for each of the states. In these optimal protocols, $P(x,y,c|a,b)=0$ for many combinations of $x,y,c,a,b$, meaning we only need to estimate fewer probabilities in the experiment. In general, we need far fewer probability combinations to certify the entanglement dimension of a target state than necessary using state tomography, as can be seen in Table~\ref{table_results}. For example,  there are only eight probability combinations needed to certify the entanglement dimension of the four-dimensional maximally entangled state $\ket{\Psi_{4}}$. This is a big improvement over both the 256 probability combinations necessary for its tomographic reconstruction and the $2d^2=32$ combinations necessary for fidelity-based entanglement dimension certification~\cite{Bavaresco2018NP}.

The theoretical and experimental values of $1-\probi-\probii$ are presented in Table \ref{table_results}. If $1-\probi-\probii>0$, the state is certified to possess $(D + 1)$-dimensional entanglement. However, we cannot get an valid conclusion (not applicable), if $1-\probi-\probii=0$. The slight difference between the experimental values and the theoretical curves may come from the differences of the noise profile: the theoretical curves are obtained by assuming the noise to be perfectly depolarizing, while the noise is added through an uncorrelated external light source in the experiment and it might have a less-than-ideal noise profile. The counting rate of the entanglement source is about $1250/s$ and all data points are collected for 20s. The statistical errors are estimated using the Monte Carlo method.

As one can appreciate, we can certify the entanglement dimension for each target state from the observation that $\probi+\probii^{\text{exp}}<1$ for all of them. The measurement bases and distributions $P(x,y,c|a,b)$ used in the experiment can be found in Appendix~\ref{app:locc} and~\ref{app:locc_prob}.

\textit{Conclusions}---
Qubits and qudits are typically made by ignoring unused degrees of freedom in physical systems. When noise inevitably affects them, unfaithful entanglement may become unavoidable in experiments. This signifies the urgent need to identify experiment-friendly entanglement detection protocols which can also certify unfaithful entanglement. In the short time since the discovery of unfaithful entanglement, advances have already been made to study its structure and detection~\cite{Guehne2021,Lo2020}. Compared to these results, our method does not specifically target unfaithful states and it is conceived to be experiment-friendly. We recast the entanglement detection problem into a quantum preparation game~\cite{WeilenmannPrepGames} and extract from the solution of the resulting optimization problem protocols capable of certifying the entanglement dimension of bipartite high-dimensional states. By incorporating an experimental technique which can produce arbitrary high-dimensional mixed states with high fidelity, we test our protocols for 4-dimensional unfaithful states. Despite having as much as $80\%$ of noise in one state, we are able to obtain close agreements between experimental and theoretical values. Our work lays the foundation for the efficient generation, detection and application~\cite{toth2018quantum,Nguyen2020} of high-dimensional entangled states.

\noindent\textit{Acknowledgments}---This work was supported by the National Key R\&D Program of China (Nos.\ 2018YFA0306703, 2021YFE0113100, 2017YFA0304100), National Natural Science  Foundation of China (Nos. 11774335, 11734015, 11874345, 11821404, 11904357,12174367), the Key Research Program of Frontier Sciences, CAS (No.\ QYZDY-SSW-SLH003), Science Foundation of the CAS (ZDRW-XH-2019-1), the Fundamental Research Funds for the Central Universities, USTC Tang Scholarship, Science and Technological Fund of Anhui Province for Outstanding Youth (2008085J02). M.N., M.W. and E.A.A. were supported by the Austrian Science Fund (FWF) stand-alone project P 30947. X.G. acknowledges the support of Austrian Academy of Sciences (\"OAW) and Joint Center for Extreme Photonics (JCEP).

\begin{appendix}

\section{Hierarchy of SDPs for high-dimensional bipartite entanglement detection}\label{app:theory}

In order to find optimal $1$-way LOCC protocols for high-dimensional entanglement detection, we invoke the general method described in~\cite{WeilenmannPrepGames} for optimizations over $1$-shot quantum preparation games. If we wish to conclude that our source can prepare bipartite states outside $S_D$ with sufficient statistical certainty, we need to repeat our $1$-shot test $n$ times. If the source is limited to only produce states in $S_D$, the maximum probability to observe $\texttt{C}$ at least $v$ times is~\cite{Elkouss2016}:

\begin{equation}
p(v, n)\equiv\sum_{k=v}^n \left(\begin{array}{c}n\\k\end{array}\right)\probi^k(1-\probi)^{n-k}.
\label{p-value}
\end{equation}
\noindent In any $n$-round experiment where the outcome $\texttt{C}$ occurs $v$ times, the quantity $p(v, n)$ can be regarded as the observed $p$-value of the null hypothesis that the entanglement source can only produce states in $S_D$.

On the other hand, if the source is actually distributing the target state $\rhot{AB}$ in each experimental round, the average observed $p$-value after $n$ rounds is upper bounded by~\cite{Araujo_2020} 
\begin{equation}
\left[1-(1-\probi-\probii)^2\right]^n,
\label{mateus_form}
\end{equation}
\noindent which motivates us to search for $1$-shot $1$-way LOCC protocols that minimize the sum $\probi+\probii$.

Now consider the following optimization problem:
\begin{align}
\underset{M,\probi,\probii}{\text{Minimize}}\quad &\probi+\probii\nonumber\\
\text{subject to} \quad &\tr(M_{\texttt{U}}\rho)=\probii,\nonumber\\
&\probi\id_{AB}-M_{\texttt{C}}\in S_D^*\nonumber\\
&M\in {\cal M}_{1-LOCC}.
\label{program}
\end{align}
\noindent Here $M=(M_{\texttt{C}}, M_{\texttt{U}})$ is a dichotomic positive operator valued measure (POVM) and ${\cal M}_{1-LOCC}$ denotes the set of 1-way LOCC POVMs achievable with measurements $\{N^x\}_{x}$. $S_D^*$ denotes the dual of the set $S_D$, i.e., ${S_D^*=\{W:\tr(W\rho)\geq 0,\forall\rho\in S_D\}}$. From the definition of dual, the second condition in (\ref{program}) is equivalent to the condition $\tr(\sigma M_C)\leq \probi$, for $\sigma\in S_D$.

In order to solve problem (\ref{program}), we need to find a simple description for the sets $S_D^*$, ${\cal M}_{1-LOCC}$. 
We start with the latter set. As shown in~\cite{WeilenmannPrepGames}, any 1-way LOCC protocol with local measurements $\{N^x\}_x$ can be modeled through a set of conditional probability distributions. More specifically, ${\cal M}_{1-LOCC}$ consists of all pairs of operators $(M_{\texttt{C}}, M_{\texttt{U}})$ for which there exists a distribution $P(x,y, c|a,b)$, with $c\in\{\texttt{C}, \texttt{U}\}$, satisfying 
\begin{align}
&P(x,y, c|a,b)\geq 0, &&\sum_c P(x,y, c|a,b)=P(x,y|a), \nonumber\\
&\sum_y P(x,y|a)=P(x), &&\sum_x P(x)=1,
\label{const}
\end{align}
\noindent and such that
\begin{align}
&M_{\texttt{C}}=M_{\texttt{C}}(P) =\sum_{x,y,a,b}P\left(x,y,c=\texttt{C}|a,b\right)N_{a|x}\otimes N_{b|y},\nonumber\\
&M_{\texttt{U}}=M_{\texttt{U}}(P) =\sum_{x,y,a,b}P\left(x,y,c=\texttt{U}|a,b\right)N_{a|x}\otimes N_{b|y}.
\label{corr}
\end{align}
\noindent See~\cite{WeilenmannPrepGames} for a prescription to derive, given any such object $P(x,y, c|a,b)$, the corresponding 1-way LOCC protocol.

Let us now focus on set $S^*_D$. Consider the ancillary systems $A',B'$, with Hilbert space dimension $D$, and call $S$ the set of states in $AA'B'B$, separable with respect to the bipartition $AA'|B'B$. In \cite{Mirjam2020}, it is proven that $S_D$ is equivalent to the set
\begin{equation}
\left\{\sigma_{AB}: \exists \omega_{AA'B'B}\in S, \mbox{ s.t. } \tr(\omega)=D\tr(\sigma), \Pi_D^\dagger\omega\Pi_D =\sigma\right\},
\end{equation}

\noindent where $\Pi =\id_{A}\otimes\ket{\psi_D^+}_{A'B'}\otimes\id_B$, and $\ket{\psi_D^+}=\sum_{i=1}^D\ket{i}\ket{i}$ is the maximally entangled state in dimension $\C^D\times\C^D$. Then $S^*_D$ can be verified to be the set of operators $W_{AB}$ such that, for some $\lambda\in \R, M\in S^*$,
\begin{equation}
\Pi_D W\Pi_D^\dagger= M +\lambda\left(\Pi_D\Pi_D^\dagger-\frac{\id}{D}\right).
\label{dual}
\end{equation}
\noindent Here $S^*$ denotes the dual of $S$, i.e., the set of witnesses for bipartite entanglement.

\begin{figure}[htbp!]
\begin{center}
\includegraphics [width=0.9\columnwidth]{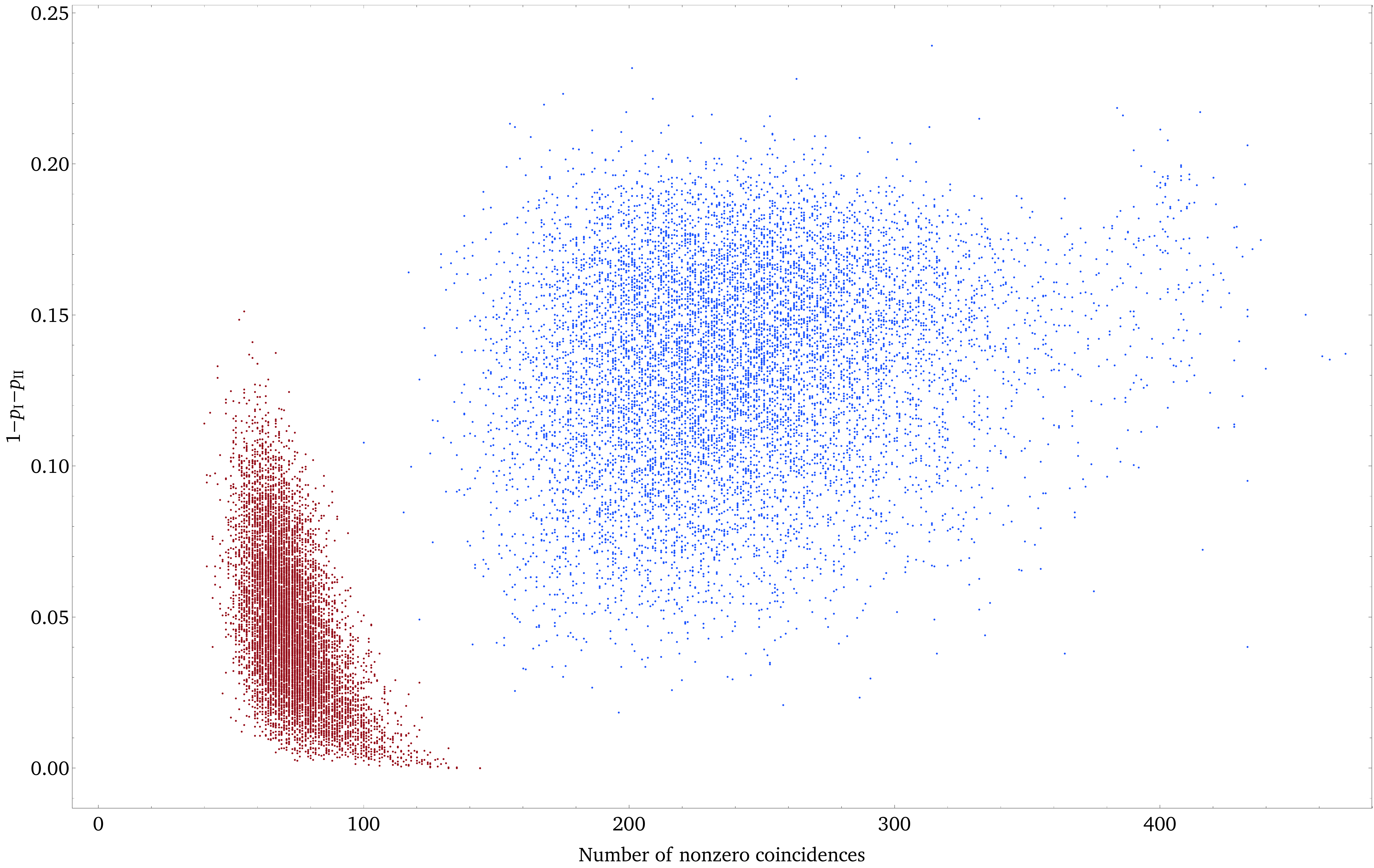}
\end{center}
\caption{Benchmarking the entanglement detection protocols found by our method: the number of nonzero probabilities vs. $1-\probi-\probii$ for 20000 random pure bipartite 4-dimensional states with Schmidt rank 2, using 3 (red) and 13 (blue) measurement settings.}
\label{fig:benchmark}
\end{figure}

Unfortunately, characterizing either set $S$ or $S^*$ is a hard problem \cite{NP_sep, strong_NP_sep}. Note, however, that, as long as $\probi\id_{AB}-M_{\texttt{C}}$ is an element of $S_D^*$, the protocol will be certified to have type-I error $\probi$ or lower. Hence we can replace $S_D^*$  in the second condition of problem \eqref{program} by any subset thereof.
  
The dual of the Doherty-Parrilo-Spedalieri hierarchy for entanglement detection provides an infinite hierarchy of semidefinite programming ans\"atze for $S^*$ that converges to $S^*$ asymptotically, see~\cite{WeilenmannPrepGames} for an explicit SDP description. A simple subset $\bar{S}^*_D$ of $S^*_D$ is obtained by requiring $M$ in decomposition (\ref{dual}) to belong to the first set of this hierarchy: the set of all operators $\Lambda_{AA'B'B}$ that can be written as $\Lambda^1 + (\Lambda^2)^{T_{BB'}}$, for some positive semidefinite matrices $\Lambda^1, \Lambda^2\geq 0$ and where ${T_{BB'}}$ denotes the partial transpose with respect to systems $BB'$.

Let us denote conditions \eqref{const} by ${\cal P}$. Putting everything together, we find that, in order to find 1-way LOCC protocols satisfying conditions 1, 2 of the main text, 
it is enough to solve the problem:
\begin{align}
\underset{M,P,\probi,\probii}{\text{Minimize}}\quad &\probi+\probii\nonumber\\
\text{subject to}\quad &\tr(M_{\texttt{U}}\rho)=p_{II},\nonumber\\
&p_{I}\id_{AB}-M_{\texttt{C}}\in \bar{S}_D^*\nonumber\\
&M_{\texttt{C}}=M_{\texttt{C}}(P), M_{\texttt{U}}=M_{\texttt{U}}(P),\nonumber\\
&P\in{\cal P}.
\end{align}
\noindent This is equivalent to problem (2) in the main text.

Even though the SDP above is valid for $d$ and $D$, in practice a number of parameters can affect the performance of the SDP solver. In addition to $d$ and $D$, the number of measurements and whether there are complex numbers in the state/measurements also contribute to the complexity of the SDP. As a rule of thumb, when $d^2D^2\approx 20$ the SDP can be solved in less than a second while $d^2D^2\approx 500$ takes a few minutes.

\begin{figure}[htbp!]
\begin{center}
\includegraphics [width=0.9\columnwidth]{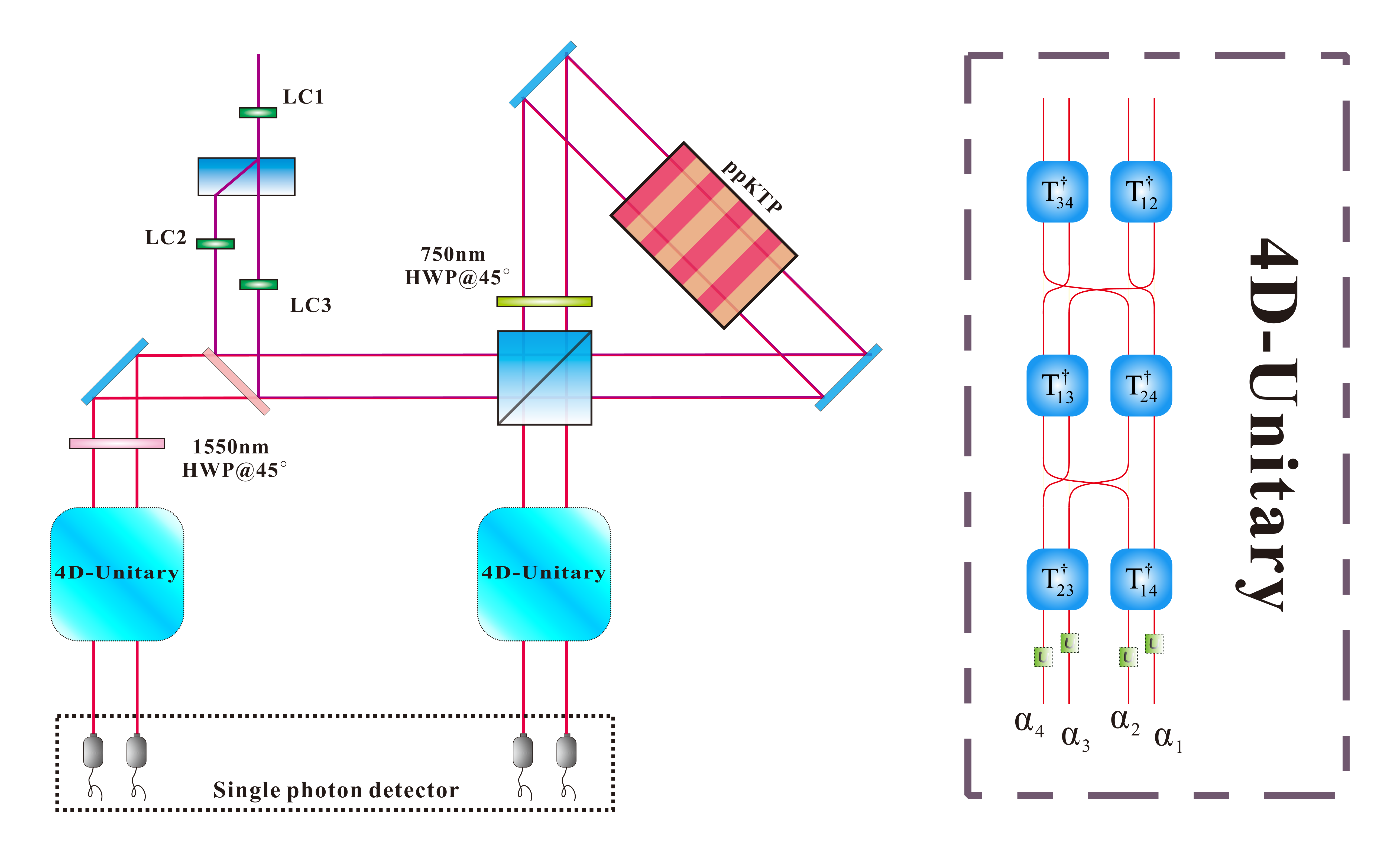}
\end{center}
\caption{Experimental scheme for the preparation of arbitrary four-dimensional two-particle state.
By adjusting the voltage loaded on LC1-LC3, state $|\psi\rangle= \lambda_{0}\left|0\right\rangle\left|0\right\rangle+\lambda_{1}\left|1\right\rangle\left|1\right\rangle+\lambda_{2}\left|2\right\rangle\left|2\right\rangle+\lambda_{3}\left|3\right\rangle\left|3\right\rangle$ ($\lambda_{0}^2+\lambda_{1}^2+\lambda_{2}^2+\lambda_{3}^2=1$) can be prepared. 
The four-dimensional pure state $|\psi\rangle=\sum_{j k} a_{j k}|j\rangle|k\rangle$ can be prepared by a local four-dimensional (4D) unitary. 
The 4D unitary operation shown in the dashed box consists of two-dimensional subspace operations and path exchanges. Finally, the target mixed state is prepared by switching the electro-optic modulation elements between these two pure states rapidly.  }
\label{fig:4D-unitary}
\end{figure}

The ability to compute the optimal entanglement detection protocol given any state and a set of measurement settings allows us to conduct a random benchmark of our method. The target states are generated randomly using the toolkit \texttt{QETLAB}~\cite{qetlab} as 4-dimensional bipartite pure states with fixed Schmidt rank 2. Two different sets of measurement settings are used: one set is used in our experiment and given by (\ref{fourier3-1})--(\ref{fourier3-3}), while the other set is generated from the eigenvectors of the 15 generators of $\mathfrak{su}(4)$. For each set of measurement settings we generated 10000 random target states and computed $1-\probi-\probii$, which indicates the robustness of the optimal protocol, and the number of nonzero probabilities that needed to be estimated in the experimental implementation. The result is shown in Fig.~\ref{fig:benchmark}. A clear trade-off can be seen: if we use more measurement setting, resulting in more probabilities, on average $1-\probi-\probii$ will be higher. On the other hand, with fewer settings, the value of $1-\probi-\probii$ on average increases when the number of probabilities decrease, indicating the experimental efficiency of optimal protocols found by our method. We expect similar plots in higher dimensions.

\section{Experimental preparation of arbitrary two-article high-dimensional mixed states.}

In this part, we are going to explain our experimental setup to prepare arbitrary two-particle high-dimensional mixed states, 
which can be decomposed into incoherent mixtures of pure states, such as $\rho=\sum_{i}P_{i}|\psi_{i}\rangle\langle\psi_{i}|$.
There are two steps during the preparation process. The first step is to prepare the two-particle pure states. The second step is to switch the electro-optic modulation elements quickly between these pure states to get the target mixed states.

According to the Schmidt decomposition, any two-particle pure state $|\psi\rangle=\sum_{j k} a_{j k}|j\rangle|k\rangle$ can be obtained by a state $|\psi\rangle=\sum_{i} \lambda_{i}\left|i_{A}\right\rangle\left|i_{B}\right\rangle$ with local unitary operations.
The transformation relation of its orthogonal basis is $\left|i_{A}\right\rangle \equiv \sum_{j} u_{j i}|j\rangle,\left|i_{B}\right\rangle \equiv \sum_{k} v_{i k}|k\rangle$. 
As an example, a four-dimensional two-particle state is prepared in our experiment, as shown in Fig. \ref{fig:4D-unitary}. 
By adjusting the voltages of LC1-LC3, one can prepare the quantum state $|\psi\rangle=\lambda_{0}\left|0\right\rangle\left|0\right\rangle+\lambda_{1}\left|1\right\rangle\left|1\right\rangle+\lambda_{2}\left|2\right\rangle\left|2\right\rangle+\lambda_{3}\left|3\right\rangle\left|3\right\rangle$ ($\lambda_{0}^2+\lambda_{1}^2+\lambda_{2}^2+\lambda_{3}^2=1$). 
To prepare the target two-particle pure quantum state, an arbitrary four-dimensional (4D) unitary shown in dashed box is necessary, which consists of six two-dimensional subspaces unitary operations together with some path-changes \cite{PhysRevLett.73.58}. 
By controlling the liquid crystals LC1 and LC2 in the six two-dimensional subspaces, one can realise the four-dimensional  unitary exchange in path.
This method is used for arbitrary high dimensions.
Then the preparation of the target mixed state is completed by adjusting the preparation time of each pure state.

\section{$1$-way LOCC protocols for entanglement detection}\label{app:locc}
In this section, we list the measurement settings and protocols used in the main text to certify the entanglement dimensions of our target states. The protocol for $\rhot{UNF}^p$ uses the three measurement settings below:

\begin{align}
N_{1|1}&=
\frac{\text{$|$0$\rangle $}}{\sqrt{3}}+\frac{1}{6} \left(-\sqrt{3}-3 i\right) \text{$|$1$\rangle $}+\frac{1}{6} \left(-\sqrt{3}+3 i\right) \text{$|$2$\rangle $},\nonumber\\
N_{2|1}&=\frac{\text{$|$0$\rangle $}}{\sqrt{3}}+\frac{1}{6} \left(-\sqrt{3}+3 i\right) \text{$|$1$\rangle $}+\frac{1}{6} \left(-\sqrt{3}-3 i\right) \text{$|$2$\rangle $},\nonumber\\
N_{3|1}&=\frac{\text{$|$0$\rangle $}}{\sqrt{3}}+\frac{\text{$|$1$\rangle $}}{\sqrt{3}}+\frac{\text{$|$2$\rangle $}}{\sqrt{3}},\, N_{4|1}=\ket{3}.\label{fourier3-1}\\
\vspace{1em}\nonumber\\
N_{1|2}&=\ket{2},\, N_{2|2}=\ket{1},\, N_{3|2}=\ket{3},\, N_{4|2}=\ket{0}.\label{fourier3-2}\\
\vspace{1em}\nonumber\\
N_{1|3}&=\frac{\text{$|$3$\rangle $}}{\sqrt{2}}-\frac{i \text{$|$2$\rangle $}}{\sqrt{2}},\, N_{2|3}=\frac{\text{$|$1$\rangle $}}{\sqrt{2}}+\frac{i \text{$|$0$\rangle $}}{\sqrt{2}},\nonumber\\
N_{3|3}&=\frac{\text{$|$3$\rangle $}}{\sqrt{2}}+\frac{i \text{$|$2$\rangle $}}{\sqrt{2}},\, N_{4|3}=\frac{\text{$|$1$\rangle $}}{\sqrt{2}}-\frac{i \text{$|$0$\rangle $}}{\sqrt{2}}.\label{fourier3-3}
\end{align}

The protocols for $\rhot{ISO2}^p$ use these three measurement settings:

\begin{align}
N_{1|1}&=\frac{-\ket{0}+\ket{1}}{\sqrt{2}},\, N_{2|1}=\ket{3},\, N_{3|1}=\ket{2},\, N_{4|1}=\frac{\ket{0}+\ket{1}}{\sqrt{2}}.\\
\vspace{1em}\nonumber\\
N_{1|2}&=\ket{3},\, N_{2|2}=\ket{2},\, N_{3|2}=\frac{-i\ket{0}+\ket{1}}{\sqrt{2}},\, N_{4|2}=\frac{i\ket{0}+\ket{1}}{\sqrt{2}}.\\
\vspace{1em}\nonumber\\
N_{1|3}&=\ket{3},\, N_{2|3}=\ket{2},\, N_{3|3}=\ket{1},\, N_{4|3}=\ket{0}.
\end{align}

There are three measurement settings needed for the protocols for the state $\Psi_{4}$, as follows:

\begin{align}
N_{1|1}&=\ket{0},\nonumber\\
N_{2|1}&=\frac{1}{\sqrt{3}}\ket{1}+\frac{1}{6} \left(-\sqrt{3}-3 i\right) \ket{2}+\frac{1}{6} \left(-\sqrt{3}+3 i\right)\ket{3},\, \nonumber\\
N_{3|1}&=\frac{1}{\sqrt{3}}\ket{1}+\frac{1}{6} \left(-\sqrt{3}+3 i\right) \ket{2}+\frac{1}{6} \left(-\sqrt{3}-3 i\right)\ket{3},\, \nonumber\\
N_{4|1}&=\frac{1}{\sqrt{3}}\ket{1}+\frac{\ket{2}}{\sqrt{3}}+\frac{\ket{3}}{\sqrt{3}},
\end{align}

\begin{align}
N_{1|1}&=\frac{1}{\sqrt{3}}\ket{0}+\frac{1}{6} \left(-\sqrt{3}-3 i\right) \ket{2}+\frac{1}{6} \left(-\sqrt{3}+3 i\right)\ket{3},\nonumber\\
N_{2|1}&=\ket{1}\, \nonumber\\
N_{3|1}&=\frac{1}{\sqrt{3}}\ket{0}+\frac{1}{6} \left(-\sqrt{3}+3 i\right) \ket{2}+\frac{1}{6} \left(-\sqrt{3}-3 i\right)\ket{3},\, \nonumber\\
N_{4|1}&=\frac{1}{\sqrt{3}}\ket{0}+\frac{\ket{2}}{\sqrt{3}}+\frac{\ket{3}}{\sqrt{3}},
\end{align}

As explained in the main text, the optimum protocol in the form of distributions $P(x,y,c|a,b)$ is part of the solution to problem (2) in the main text found by the SDP solver. Using the 3 measurement settings above, we obtain different distributions for different target states. For some values of $x,y,a,b$, $P(x,y,c|a,b)$ in an optimal protocol will be zero. As a result, we will only list the nonzero values of $P(x,y,c|a,b)$. The values use in different measurement protocols are listed in Appendix~\ref{app:locc_prob}.

\section{Additional experimental details}\label{app:exp}
\begin{figure}[htbp!]
	\centering
	\vspace{0.6cm}
	\includegraphics [width=0.6\columnwidth]{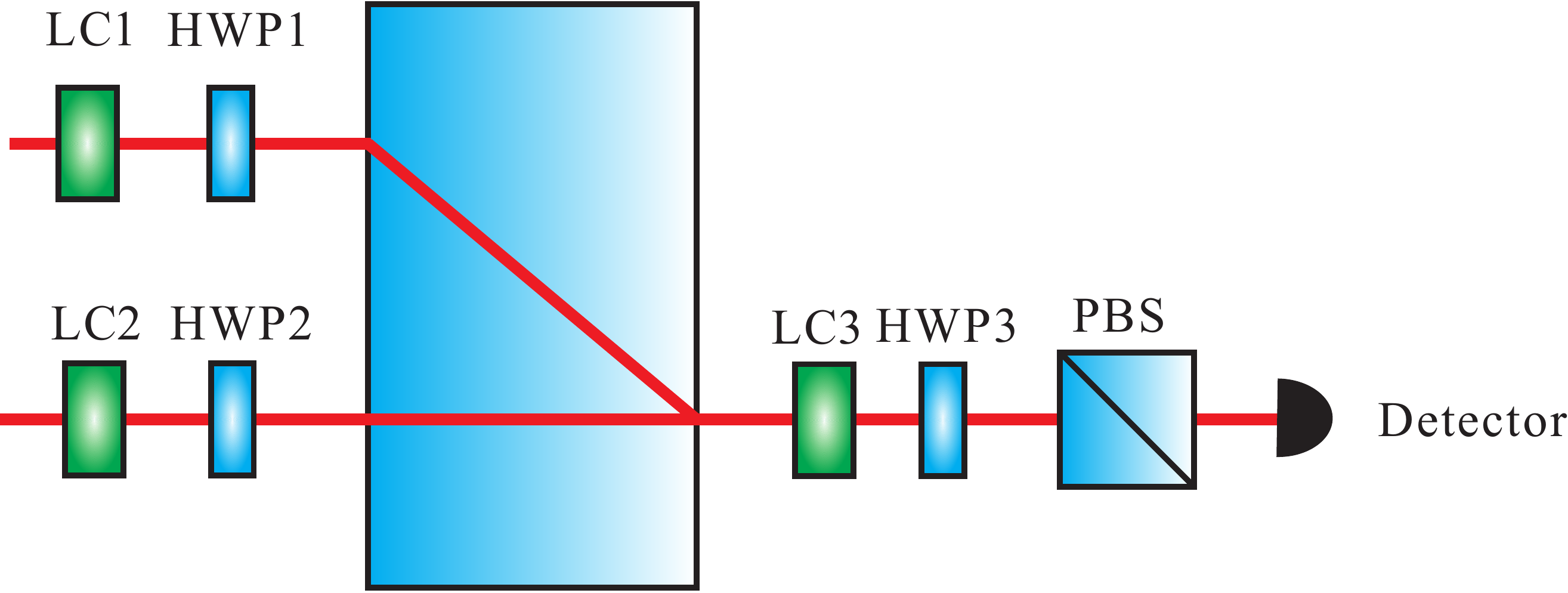}
	\caption{Detailed setup for implementing projective measurements. All HWPs are set at $22.5^{\circ}$. By adjusting the voltages of LCs(1-3), one can obtain the projective measurement of $(|0\rangle+e^{i\varphi_{1}}|1\rangle+e^{i\varphi_{3}}(|2\rangle+e^{i\varphi_{2}}|3\rangle))/2$. By adjusting the angle of HWP1-HWP3 and the voltage of LC1-LC3, arbitrary four-dimensional measurement can be realized.}
	\label{fig:projector}
\end{figure}

Our LOCC protocols require us to implement many different projective measurements. This is achieved by adjusting the voltages of the LCs and angles of HWPs on Alice's and Bob's local measurement setups. An illustrative example is given in Fig.~\ref{fig:projector}.

\begin{figure}[htbp!]
\begin{center}
\includegraphics [width=0.6\columnwidth]{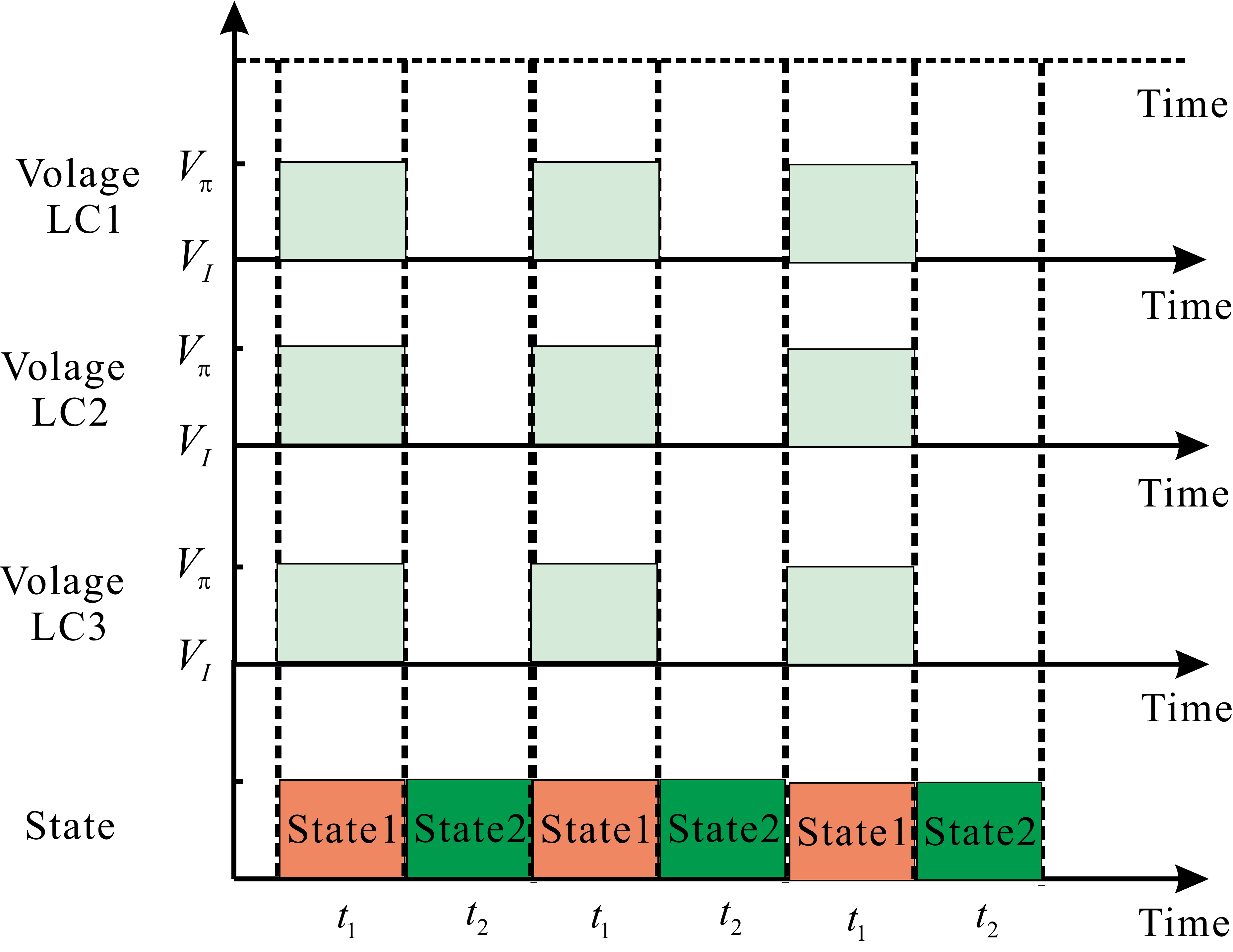}
\end{center}
\caption{Voltage timings of LCs for preparing the two pure states needed to obtain $\rhot{UNF}$. The difference between $t_1$ and $t_2$ corresponds to the mixing ratio. To make $\rhot{UNF}$, we simply make $t_1=t_2=1s$.}
\label{fig:mixrho}
\end{figure}

As described in the main text, we obtained the state $\rhot{UNF}^{p}$ by first preparing two pure states $\ket{\Psi_{3}}$ and $\frac{1}{\sqrt{2}}(\ket{23}+\ket{32})$. By switching the voltages applied to LC1-LC3 as shown in Fig.~\ref{fig:mixrho}, we mix these two pure states into the state $\rhot{UNF}^{p}$.

The states $\ket{\Psi_{3}}$ and $\frac{1}{\sqrt{2}}(\ket{23}+\ket{32})$ can be generated by applying voltages $V_I$ and $V_{\pi}$ to the first three LCs, respectively, while setting their angles at $27.4^{\circ}$, $22.5^{\circ}$, and $45^{\circ}$. The angles of the first three HWPs are set at $27.4^{\circ}$, $22.5^{\circ}$, and $45^{\circ}$. By rapidly switching the voltage of the first three LCs between $V_I$ and $V_{\pi}$, we obtain the state $\rhot{UNF}^{p}$.

The states $\rhot{ISO2}^p$ are prepared by mixing a 2-dimensional Bell state with 4-dimensional noise. 
First, we prepare the Bell state $\frac{1}{ \sqrt{2}}(\ket{00}+\ket{11})$, then noise is added by coupling two independent light sources to the measurement setup. The light sources, which are variable-intensity LEDs, are positioned before two couplers to allow independent adjustment of the noise entering each detector. The light is coupled to each coupler by diffuse reflection and the total number of photons entering the coupler can be
controlled by changing the brightness of the LEDs. Using the brightness of the entanglement source (coincidence counts) and the added noise (coincidence noise), we then estimate the loaded white noise count on a single channel. The coincidence count $C$ can be computed from the (empirical) formula of random coincidences:
\begin{equation}
C=2\times S_{A}\times S_{B} \times \tau, 
\end{equation}
where $\tau$ is the coincidence window and $S_A, \, S_B$ are single-channel counts in Alice's and Bob's couplers.
In our experiment, $\tau=5\times 10^{-9}s$. We can change the amount of noise by adjusting the intensity of the light sources.

As explained in the main text, we performed the tomographic reconstruction of the state generated by our setup to certify its unfaithfulness. We here selected $\rhot{UNF}^{0.2}$, $\rhot{UNF}^{0.6}$, $\rhot{SIO2}^{0.6}$, and $\rhot{SIO2}^{0.8}$ for quantum state tomography. The fidelities of the reconstructed states with respect to the ideal states are $0.989\pm0.001$, $0.990\pm0.001$, $0.993\pm0.004$, and $0.996\pm0.002$, respectively.

\section{Probabilities for 1-Way LOCC protocols}\label{app:locc_prob}

\begin{table}[htbp!]
\scriptsize
\centering
\begin{tabular}{|c|c|c|c|c|c|c|c|c|c|}
\hline
$x$&$y$&$a$&$b$&$P(x,y,\texttt{C}|a,b)$&$x$&$y$&$a$&$b$&$P(x,y,\texttt{C}|a,b)$\\
\hline
1 & 1 & 1 & 2 & 0.459 & 2 & 2 & 2 & 2 & 0.390 \\
\hline
1 & 1 & 1 & 4 & 0.229 & 2 & 2 & 3 & 1 & 0.277 \\
\hline
1 & 1 & 2 & 1 & 0.459 & 2 & 2 & 4 & 4 & 0.390 \\
\hline
1 & 1 & 2 & 4 & 0.229 & 2 & 3 & 1 & 1 & 0.195 \\
\hline
1 & 1 & 3 & 3 & 0.459 & 2 & 3 & 1 & 3 & 0.195 \\
\hline
1 & 1 & 3 & 4 & 0.223 & 3 & 1 & 2 & 1 & 0.009 \\
\hline
1 & 1 & 4 & 1 & 0.003 & 3 & 1 & 2 & 3 & 0.003 \\
\hline
1 & 1 & 4 & 2 & 0.003 & 3 & 1 & 4 & 2 & 0.009 \\
\hline
1 & 2 & 4 & 1 & 0.336 & 3 & 1 & 4 & 3 & 0.003 \\
\hline
2 & 1 & 3 & 1 & 0.003 & 3 & 3 & 1 & 1 & 0.151 \\
\hline
2 & 1 & 3 & 2 & 0.003 & 3 & 3 & 2 & 4 & 0.104 \\
\hline
2 & 2 & 1 & 1 & 0.195 & 3 & 3 & 3 & 3 & 0.151 \\
\hline
2 & 2 & 1 & 3 & 0.195 & 3 & 3 & 4 & 2 & 0.104 \\
\hline
\end{tabular}
\caption{The LOCC protocol for target state $\rhot{UNF3}^{0.4}$,$\rhot{UNF3}^{0.6}$.}
\label{table:rhounf3}

\end{table}

\begin{table}[htbp!]
\scriptsize
\centering
\begin{tabular}{|c|c|c|c|c|c|c|c|c|c|}
\hline
$x$&$y$&$a$&$b$&$P(x,y,\texttt{C}|a,b)$&$x$&$y$&$a$&$b$&$P(x,y,\texttt{C}|a,b)$\\
\hline
1 & 1 & 1 & 2 & 0.500 & 2 & 1 & 3 & 2 & 0.087 \\
\hline
1 & 1 & 1 & 4 & 0.223 & 2 & 1 & 3 & 3 & 0.088 \\
\hline
1 & 1 & 2 & 1 & 0.500 & 2 & 1 & 3 & 4 & 0.088 \\
\hline
1 & 1 & 2 & 4 & 0.223 & 2 & 2 & 1 & 1 & 0.299 \\
\hline
1 & 1 & 3 & 3 & 0.500 & 2 & 2 & 1 & 3 & 0.113 \\
\hline
1 & 1 & 3 & 4 & 0.223 & 2 & 2 & 2 & 2 & 0.500 \\
\hline
1 & 1 & 4 & 1 & 0.087 & 2 & 2 & 2 & 3 & 0.301 \\
\hline
1 & 1 & 4 & 2 & 0.087 & 2 & 2 & 3 & 1 & 0.088 \\
\hline
1 & 1 & 4 & 3 & 0.087 & 2 & 2 & 3 & 2 & 0.087 \\
\hline
1 & 1 & 4 & 4 & 0.087 & 2 & 2 & 3 & 3 & 0.088 \\
\hline
1 & 2 & 4 & 1 & 0.088 & 2 & 2 & 3 & 4 & 0.087 \\
\hline
1 & 2 & 4 & 2 & 0.088 & 2 & 2 & 4 & 3 & 0.301 \\
\hline
1 & 2 & 4 & 3 & 0.087 & 2 & 2 & 4 & 4 & 0.500 \\
\hline
1 & 2 & 4 & 4 & 0.088 & 2 & 3 & 1 & 1 & 0.201 \\
\hline
1 & 3 & 4 & 1 & 0.088 & 2 & 3 & 1 & 3 & 0.201 \\
\hline
1 & 3 & 4 & 2 & 0.087 & 2 & 3 & 3 & 1 & 0.088 \\
\hline
1 & 3 & 4 & 3 & 0.088 & 2 & 3 & 3 & 2 & 0.087 \\
\hline
1 & 3 & 4 & 4 & 0.087 & 2 & 3 & 3 & 3 & 0.088 \\
\hline
2 & 1 & 3 & 1 & 0.087 & 2 & 3 & 3 & 4 & 0.087 \\
\hline
\end{tabular}
\caption{The LOCC protocol for target state $\rhot{UNF3}^{0.00}$.}
\label{table:rhounf1}

\end{table}

\begin{table}[htbp!]
\scriptsize
\centering
\begin{tabular}{|c|c|c|c|c|c|c|c|c|c|}
\hline
$x$&$y$&$a$&$b$&$P(x,y,\texttt{C}|a,b)$&$x$&$y$&$a$&$b$&$P(x,y,\texttt{C}|a,b)$\\
\hline
1 & 1 & 1 & 2 & 0.485 & 2 & 2 & 1 & 1 & 0.238 \\
\hline
1 & 1 & 1 & 4 & 0.214 & 2 & 2 & 1 & 3 & 0.238 \\
\hline
1 & 1 & 2 & 1 & 0.485 & 2 & 2 & 2 & 2 & 0.471 \\
\hline
1 & 1 & 2 & 4 & 0.214 & 2 & 2 & 3 & 1 & 0.337 \\
\hline
1 & 1 & 3 & 3 & 0.485 & 2 & 2 & 4 & 4 & 0.471 \\
\hline
1 & 1 & 3 & 4 & 0.213 & 2 & 3 & 1 & 1 & 0.233 \\
\hline
1 & 1 & 4 & 1 & 0.001 & 2 & 3 & 1 & 3 & 0.233 \\
\hline
1 & 1 & 4 & 2 & 0.001 & 3 & 3 & 1 & 1 & 0.043 \\
\hline
1 & 2 & 4 & 1 & 0.347 & 3 & 3 & 2 & 4 & 0.022 \\
\hline
2 & 1 & 3 & 1 & 0.001 & 3 & 3 & 3 & 3 & 0.043 \\
\hline
2 & 1 & 3 & 2 & 0.001 & 3 & 3 & 4 & 2 & 0.022 \\
\hline
\end{tabular}
\caption{The LOCC protocol for target state $\rhot{UNF3}^{0.2}$.}
\label{table:rhounf2}

\end{table}

\begin{table}[htbp!]
\scriptsize
\centering
\begin{tabular}{|c|c|c|c|c|c|c|c|c|c|}
\hline
$x$&$y$&$a$&$b$&$P(x,y,\texttt{C}|a,b)$&$x$&$y$&$a$&$b$&$P(x,y,\texttt{C}|a,b)$\\
\hline
1 & 1 & 1 & 2 & 0.461 & 2 & 2 & 3 & 2 & 0.000 \\
\hline
1 & 1 & 1 & 4 & 0.321 & 2 & 2 & 3 & 4 & 0.000 \\
\hline
1 & 1 & 2 & 1 & 0.461 & 2 & 2 & 4 & 3 & 0.000 \\
\hline
1 & 1 & 2 & 4 & 0.321 & 2 & 2 & 4 & 4 & 0.268 \\
\hline
1 & 1 & 3 & 3 & 0.461 & 2 & 3 & 1 & 1 & 0.136 \\
\hline
1 & 1 & 3 & 4 & 0.321 & 2 & 3 & 1 & 3 & 0.136 \\
\hline
1 & 1 & 4 & 1 & 0.000 & 2 & 3 & 3 & 2 & 0.000 \\
\hline
1 & 1 & 4 & 2 & 0.000 & 2 & 3 & 3 & 4 & 0.000 \\
\hline
1 & 1 & 4 & 3 & 0.000 & 3 & 1 & 2 & 1 & 0.002 \\
\hline
1 & 2 & 4 & 1 & 0.390 & 3 & 1 & 2 & 3 & 0.002 \\
\hline
1 & 2 & 4 & 2 & 0.000 & 3 & 1 & 2 & 4 & 0.000 \\
\hline
1 & 2 & 4 & 4 & 0.000 & 3 & 1 & 4 & 2 & 0.002 \\
\hline
1 & 3 & 4 & 2 & 0.000 & 3 & 1 & 4 & 3 & 0.002 \\
\hline
1 & 3 & 4 & 4 & 0.000 & 3 & 1 & 4 & 4 & 0.000 \\
\hline
2 & 1 & 3 & 1 & 0.000 & 3 & 2 & 2 & 2 & 0.000 \\
\hline
2 & 1 & 3 & 2 & 0.000 & 3 & 2 & 2 & 4 & 0.000 \\
\hline
2 & 1 & 3 & 3 & 0.000 & 3 & 2 & 4 & 2 & 0.000 \\
\hline
2 & 2 & 1 & 1 & 0.132 & 3 & 2 & 4 & 4 & 0.000 \\
\hline
2 & 2 & 1 & 3 & 0.132 & 3 & 3 & 1 & 1 & 0.271 \\
\hline
2 & 2 & 2 & 2 & 0.268 & 3 & 3 & 2 & 4 & 0.269 \\
\hline
2 & 2 & 2 & 3 & 0.000 & 3 & 3 & 3 & 3 & 0.271 \\
\hline
2 & 2 & 3 & 1 & 0.198 & 3 & 3 & 4 & 2 & 0.269 \\
\hline
\end{tabular}
\caption{The LOCC protocol for target state $\rhot{UNF3}^{0.8}$.}
\label{table:rhounf4}

\end{table}

\begin{table}[htbp!]
\scriptsize
\centering
\begin{tabular}{|c|c|c|c|c|c|c|c|c|c|c|c|c|c|c|}
\hline
$x$&$y$&$a$&$b$&$P(x,y,\texttt{C}|a,b)$&$x$&$y$&$a$&$b$&$P(x,y,\texttt{C}|a,b)$&$x$&$y$&$a$&$b$&$P(x,y,\texttt{C}|a,b)$\\
\hline
1 & 1 & 1 & 1 & 0.333 & 2 & 1 & 1 & 1 & 0.074 & 3 & 1 & 1 & 1 & 0.074 \\
\hline
1 & 1 & 1 & 2 & 0.222 & 2 & 1 & 1 & 2 & 0.074 & 3 & 1 & 1 & 2 & 0.074 \\
\hline
1 & 1 & 1 & 3 & 0.222 & 2 & 1 & 1 & 3 & 0.074 & 3 & 1 & 1 & 3 & 0.074 \\
\hline
1 & 1 & 2 & 1 & 0.074 & 2 & 1 & 1 & 4 & 0.074 & 3 & 1 & 1 & 4 & 0.074 \\
\hline
1 & 1 & 2 & 2 & 0.074 & 2 & 1 & 2 & 1 & 0.074 & 3 & 1 & 2 & 1 & 0.074 \\
\hline
1 & 1 & 2 & 3 & 0.074 & 2 & 1 & 2 & 2 & 0.074 & 3 & 1 & 2 & 2 & 0.074 \\
\hline
1 & 1 & 2 & 4 & 0.074 & 2 & 1 & 2 & 3 & 0.074 & 3 & 1 & 2 & 3 & 0.074 \\
\hline
1 & 1 & 3 & 1 & 0.074 & 2 & 1 & 2 & 4 & 0.074 & 3 & 1 & 2 & 4 & 0.074 \\
\hline
1 & 1 & 3 & 2 & 0.074 & 2 & 2 & 1 & 1 & 0.074 & 3 & 2 & 1 & 1 & 0.074 \\
\hline
1 & 1 & 3 & 3 & 0.074 & 2 & 2 & 1 & 2 & 0.074 & 3 & 2 & 1 & 2 & 0.074 \\
\hline
1 & 1 & 3 & 4 & 0.074 & 2 & 2 & 1 & 3 & 0.074 & 3 & 2 & 1 & 3 & 0.074 \\
\hline
1 & 1 & 4 & 2 & 0.222 & 2 & 2 & 1 & 4 & 0.074 & 3 & 2 & 1 & 4 & 0.074 \\
\hline
1 & 1 & 4 & 3 & 0.222 & 2 & 2 & 2 & 1 & 0.074 & 3 & 2 & 2 & 1 & 0.074 \\
\hline
1 & 1 & 4 & 4 & 0.333 & 2 & 2 & 2 & 2 & 0.074 & 3 & 2 & 2 & 2 & 0.074 \\
\hline
1 & 2 & 2 & 1 & 0.074 & 2 & 2 & 2 & 3 & 0.074 & 3 & 2 & 2 & 3 & 0.074 \\
\hline
1 & 2 & 2 & 2 & 0.074 & 2 & 2 & 2 & 4 & 0.074 & 3 & 2 & 2 & 4 & 0.074 \\
\hline
1 & 2 & 2 & 3 & 0.074 & 2 & 2 & 3 & 1 & 0.222 & 3 & 3 & 1 & 1 & 0.074 \\
\hline
1 & 2 & 2 & 4 & 0.074 & 2 & 2 & 3 & 2 & 0.222 & 3 & 3 & 1 & 2 & 0.074 \\
\hline
1 & 2 & 3 & 1 & 0.074 & 2 & 2 & 3 & 4 & 0.333 & 3 & 3 & 1 & 3 & 0.074 \\
\hline
1 & 2 & 3 & 2 & 0.074 & 2 & 2 & 4 & 1 & 0.222 & 3 & 3 & 1 & 4 & 0.074 \\
\hline
1 & 2 & 3 & 3 & 0.074 & 2 & 2 & 4 & 2 & 0.222 & 3 & 3 & 2 & 1 & 0.074 \\
\hline
1 & 2 & 3 & 4 & 0.074 & 2 & 2 & 4 & 3 & 0.333 & 3 & 3 & 2 & 2 & 0.074 \\
\hline
1 & 3 & 2 & 1 & 0.074 & 2 & 3 & 1 & 1 & 0.074 & 3 & 3 & 2 & 3 & 0.074 \\
\hline
1 & 3 & 2 & 2 & 0.074 & 2 & 3 & 1 & 2 & 0.074 & 3 & 3 & 2 & 4 & 0.074 \\
\hline
1 & 3 & 2 & 3 & 0.074 & 2 & 3 & 1 & 3 & 0.074 & 3 & 3 & 3 & 1 & 0.222 \\
\hline
1 & 3 & 2 & 4 & 0.074 & 2 & 3 & 1 & 4 & 0.074 & 3 & 3 & 3 & 2 & 0.222 \\
\hline
1 & 3 & 3 & 1 & 0.074 & 2 & 3 & 2 & 1 & 0.074 & 3 & 3 & 3 & 3 & 0.333 \\
\hline
1 & 3 & 3 & 2 & 0.074 & 2 & 3 & 2 & 2 & 0.074 & 3 & 3 & 4 & 1 & 0.222 \\
\hline
1 & 3 & 3 & 3 & 0.074 & 2 & 3 & 2 & 3 & 0.074 & 3 & 3 & 4 & 2 & 0.222 \\
\hline
1 & 3 & 3 & 4 & 0.074 & 2 & 3 & 2 & 4 & 0.074 & 3 & 3 & 4 & 4 & 0.333 \\
\hline
\end{tabular}
\caption{The LOCC protocol for target state $\rhot{ISO2}^{0.5}$, $\rhot{ISO2}^{0.6}$, $\rhot{ISO2}^{0.7}$ and $\rhot{ISO2}^{0.8}$.}
\label{table:rhoiso2.1}

\end{table}

\begin{table}[htbp!]
\scriptsize
\centering
\begin{tabular}{|c|c|c|c|c|c|c|c|c|c|c|c|c|c|c|}
\hline
$x$&$y$&$a$&$b$&$P(x,y,\texttt{C}|a,b)$&$x$&$y$&$a$&$b$&$P(x,y,\texttt{C}|a,b)$&$x$&$y$&$a$&$b$&$P(x,y,\texttt{C}|a,b)$\\
\hline
1 & 1 & 1 & 1 & 0.056 & 2 & 1 & 1 & 1 & 0.056 & 3 & 1 & 1 & 1 & 0.056 \\
\hline
1 & 1 & 1 & 2 & 0.057 & 2 & 1 & 1 & 2 & 0.056 & 3 & 1 & 1 & 2 & 0.056 \\
\hline
1 & 1 & 1 & 3 & 0.057 & 2 & 1 & 1 & 3 & 0.056 & 3 & 1 & 1 & 3 & 0.056 \\
\hline
1 & 1 & 1 & 4 & 0.056 & 2 & 1 & 1 & 4 & 0.056 & 3 & 1 & 1 & 4 & 0.056 \\
\hline
1 & 1 & 2 & 1 & 0.056 & 2 & 1 & 2 & 1 & 0.056 & 3 & 1 & 2 & 1 & 0.056 \\
\hline
1 & 1 & 2 & 2 & 0.056 & 2 & 1 & 2 & 2 & 0.056 & 3 & 1 & 2 & 2 & 0.056 \\
\hline
1 & 1 & 2 & 3 & 0.056 & 2 & 1 & 2 & 3 & 0.056 & 3 & 1 & 2 & 3 & 0.056 \\
\hline
1 & 1 & 2 & 4 & 0.056 & 2 & 1 & 2 & 4 & 0.056 & 3 & 1 & 2 & 4 & 0.056 \\
\hline
1 & 1 & 3 & 1 & 0.056 & 2 & 1 & 3 & 1 & 0.056 & 3 & 1 & 3 & 1 & 0.056 \\
\hline
1 & 1 & 3 & 2 & 0.056 & 2 & 1 & 3 & 2 & 0.056 & 3 & 1 & 3 & 2 & 0.056 \\
\hline
1 & 1 & 3 & 3 & 0.056 & 2 & 1 & 3 & 3 & 0.056 & 3 & 1 & 3 & 3 & 0.056 \\
\hline
1 & 1 & 3 & 4 & 0.056 & 2 & 1 & 3 & 4 & 0.056 & 3 & 1 & 3 & 4 & 0.056 \\
\hline
1 & 1 & 4 & 1 & 0.056 & 2 & 1 & 4 & 1 & 0.056 & 3 & 1 & 4 & 1 & 0.056 \\
\hline
1 & 1 & 4 & 2 & 0.057 & 2 & 1 & 4 & 2 & 0.056 & 3 & 1 & 4 & 2 & 0.056 \\
\hline
1 & 1 & 4 & 3 & 0.057 & 2 & 1 & 4 & 3 & 0.056 & 3 & 1 & 4 & 3 & 0.056 \\
\hline
1 & 1 & 4 & 4 & 0.056 & 2 & 1 & 4 & 4 & 0.056 & 3 & 1 & 4 & 4 & 0.056 \\
\hline
1 & 2 & 1 & 1 & 0.056 & 2 & 2 & 1 & 1 & 0.056 & 3 & 2 & 1 & 1 & 0.056 \\
\hline
1 & 2 & 1 & 2 & 0.056 & 2 & 2 & 1 & 2 & 0.056 & 3 & 2 & 1 & 2 & 0.056 \\
\hline
1 & 2 & 1 & 3 & 0.056 & 2 & 2 & 1 & 3 & 0.056 & 3 & 2 & 1 & 3 & 0.056 \\
\hline
1 & 2 & 1 & 4 & 0.056 & 2 & 2 & 1 & 4 & 0.056 & 3 & 2 & 1 & 4 & 0.056 \\
\hline
1 & 2 & 2 & 1 & 0.056 & 2 & 2 & 2 & 1 & 0.056 & 3 & 2 & 2 & 1 & 0.056 \\
\hline
1 & 2 & 2 & 2 & 0.056 & 2 & 2 & 2 & 2 & 0.056 & 3 & 2 & 2 & 2 & 0.056 \\
\hline
1 & 2 & 2 & 3 & 0.056 & 2 & 2 & 2 & 3 & 0.056 & 3 & 2 & 2 & 3 & 0.056 \\
\hline
1 & 2 & 2 & 4 & 0.056 & 2 & 2 & 2 & 4 & 0.056 & 3 & 2 & 2 & 4 & 0.056 \\
\hline
1 & 2 & 3 & 1 & 0.056 & 2 & 2 & 3 & 1 & 0.057 & 3 & 2 & 3 & 1 & 0.056 \\
\hline
1 & 2 & 3 & 2 & 0.056 & 2 & 2 & 3 & 2 & 0.057 & 3 & 2 & 3 & 2 & 0.056 \\
\hline
1 & 2 & 3 & 3 & 0.056 & 2 & 2 & 3 & 3 & 0.056 & 3 & 2 & 3 & 3 & 0.056 \\
\hline
1 & 2 & 3 & 4 & 0.056 & 2 & 2 & 3 & 4 & 0.056 & 3 & 2 & 3 & 4 & 0.056 \\
\hline
1 & 2 & 4 & 1 & 0.056 & 2 & 2 & 4 & 1 & 0.057 & 3 & 2 & 4 & 1 & 0.056 \\
\hline
1 & 2 & 4 & 2 & 0.056 & 2 & 2 & 4 & 2 & 0.057 & 3 & 2 & 4 & 2 & 0.056 \\
\hline
1 & 2 & 4 & 3 & 0.056 & 2 & 2 & 4 & 3 & 0.056 & 3 & 2 & 4 & 3 & 0.056 \\
\hline
1 & 2 & 4 & 4 & 0.056 & 2 & 2 & 4 & 4 & 0.056 & 3 & 2 & 4 & 4 & 0.056 \\
\hline
1 & 3 & 1 & 1 & 0.056 & 2 & 3 & 1 & 1 & 0.056 & 3 & 3 & 1 & 1 & 0.056 \\
\hline
1 & 3 & 1 & 2 & 0.056 & 2 & 3 & 1 & 2 & 0.056 & 3 & 3 & 1 & 2 & 0.056 \\
\hline
1 & 3 & 1 & 3 & 0.056 & 2 & 3 & 1 & 3 & 0.056 & 3 & 3 & 1 & 3 & 0.056 \\
\hline
1 & 3 & 1 & 4 & 0.056 & 2 & 3 & 1 & 4 & 0.056 & 3 & 3 & 1 & 4 & 0.056 \\
\hline
1 & 3 & 2 & 1 & 0.056 & 2 & 3 & 2 & 1 & 0.056 & 3 & 3 & 2 & 1 & 0.056 \\
\hline
1 & 3 & 2 & 2 & 0.056 & 2 & 3 & 2 & 2 & 0.056 & 3 & 3 & 2 & 2 & 0.056 \\
\hline
1 & 3 & 2 & 3 & 0.056 & 2 & 3 & 2 & 3 & 0.056 & 3 & 3 & 2 & 3 & 0.056 \\
\hline
1 & 3 & 2 & 4 & 0.056 & 2 & 3 & 2 & 4 & 0.056 & 3 & 3 & 2 & 4 & 0.056 \\
\hline
1 & 3 & 3 & 1 & 0.056 & 2 & 3 & 3 & 1 & 0.056 & 3 & 3 & 3 & 1 & 0.057 \\
\hline
1 & 3 & 3 & 2 & 0.056 & 2 & 3 & 3 & 2 & 0.056 & 3 & 3 & 3 & 2 & 0.057 \\
\hline
1 & 3 & 3 & 3 & 0.056 & 2 & 3 & 3 & 3 & 0.056 & 3 & 3 & 3 & 3 & 0.056 \\
\hline
1 & 3 & 3 & 4 & 0.056 & 2 & 3 & 3 & 4 & 0.056 & 3 & 3 & 3 & 4 & 0.056 \\
\hline
1 & 3 & 4 & 1 & 0.056 & 2 & 3 & 4 & 1 & 0.056 & 3 & 3 & 4 & 1 & 0.057 \\
\hline
1 & 3 & 4 & 2 & 0.056 & 2 & 3 & 4 & 2 & 0.056 & 3 & 3 & 4 & 2 & 0.057 \\
\hline
1 & 3 & 4 & 3 & 0.056 & 2 & 3 & 4 & 3 & 0.056 & 3 & 3 & 4 & 3 & 0.056 \\
\hline
1 & 3 & 4 & 4 & 0.056 & 2 & 3 & 4 & 4 & 0.056 & 3 & 3 & 4 & 4 & 0.056 \\
\hline
\end{tabular}
\caption{The LOCC protocol for target state $\rhot{ISO2}^{0.9}$.}
\label{table:rhoiso2.2}

\end{table}

\begin{table*}[htbp!]
\scriptsize
\centering
\begin{tabular}{|c|c|c|c|c|c|c|c|c|c|c|c|c|c|c|}
\hline
$x$&$y$&$a$&$b$&$P(x,y,\texttt{C}|a,b)$&$x$&$y$&$a$&$b$&$P(x,y,\texttt{C}|a,b)$&$x$&$y$&$a$&$b$&$P(x,y,\texttt{C}|a,b)$\\
\hline
1 & 1 & 1 & 1 & 0.046 & 2 & 1 & 1 & 1 & 0.038 & 3 & 1 & 1 & 1 & 0.031 \\
\hline
1 & 1 & 1 & 2 & 0.046 & 2 & 1 & 1 & 2 & 0.038 & 3 & 1 & 1 & 2 & 0.031 \\
\hline
1 & 1 & 1 & 3 & 0.046 & 2 & 1 & 1 & 3 & 0.038 & 3 & 1 & 1 & 3 & 0.031 \\
\hline
1 & 1 & 1 & 4 & 0.066 & 2 & 1 & 1 & 4 & 0.095 & 3 & 1 & 1 & 4 & 0.035 \\
\hline
1 & 1 & 2 & 1 & 0.046 & 2 & 1 & 2 & 1 & 0.052 & 3 & 1 & 2 & 1 & 0.063 \\
\hline
1 & 1 & 2 & 2 & 0.046 & 2 & 1 & 2 & 2 & 0.052 & 3 & 1 & 2 & 2 & 0.063 \\
\hline
1 & 1 & 2 & 3 & 0.046 & 2 & 1 & 2 & 3 & 0.052 & 3 & 1 & 2 & 3 & 0.063 \\
\hline
1 & 1 & 2 & 4 & 0.066 & 2 & 1 & 2 & 4 & 0.047 & 3 & 1 & 2 & 4 & 0.054 \\
\hline
1 & 1 & 3 & 1 & 0.046 & 2 & 1 & 3 & 1 & 0.065 & 3 & 1 & 3 & 1 & 0.031 \\
\hline
1 & 1 & 3 & 2 & 0.046 & 2 & 1 & 3 & 2 & 0.065 & 3 & 1 & 3 & 2 & 0.031 \\
\hline
1 & 1 & 3 & 3 & 0.046 & 2 & 1 & 3 & 3 & 0.065 & 3 & 1 & 3 & 3 & 0.031 \\
\hline
1 & 1 & 3 & 4 & 0.066 & 2 & 1 & 3 & 4 & 0.019 & 3 & 1 & 3 & 4 & 0.035 \\
\hline
1 & 1 & 4 & 1 & 0.061 & 2 & 1 & 4 & 1 & 0.052 & 3 & 1 & 4 & 1 & 0.063 \\
\hline
1 & 1 & 4 & 2 & 0.061 & 2 & 1 & 4 & 2 & 0.052 & 3 & 1 & 4 & 2 & 0.063 \\
\hline
1 & 1 & 4 & 3 & 0.061 & 2 & 1 & 4 & 3 & 0.052 & 3 & 1 & 4 & 3 & 0.063 \\
\hline
1 & 1 & 4 & 4 & 0.019 & 2 & 1 & 4 & 4 & 0.047 & 3 & 1 & 4 & 4 & 0.054 \\
\hline
1 & 2 & 1 & 1 & 0.036 & 2 & 2 & 1 & 1 & 0.006 & 3 & 2 & 1 & 1 & 0.029 \\
\hline
1 & 2 & 1 & 2 & 0.051 & 2 & 2 & 1 & 2 & 0.069 & 3 & 2 & 1 & 2 & 0.032 \\
\hline
1 & 2 & 1 & 3 & 0.067 & 2 & 2 & 1 & 3 & 0.112 & 3 & 2 & 1 & 3 & 0.036 \\
\hline
1 & 2 & 1 & 4 & 0.051 & 2 & 2 & 1 & 4 & 0.069 & 3 & 2 & 1 & 4 & 0.032 \\
\hline
1 & 2 & 2 & 1 & 0.036 & 2 & 2 & 2 & 1 & 0.055 & 3 & 2 & 2 & 1 & 0.069 \\
\hline
1 & 2 & 2 & 2 & 0.051 & 2 & 2 & 2 & 2 & 0.051 & 3 & 2 & 2 & 2 & 0.061 \\
\hline
1 & 2 & 2 & 3 & 0.067 & 2 & 2 & 2 & 3 & 0.047 & 3 & 2 & 2 & 3 & 0.055 \\
\hline
1 & 2 & 2 & 4 & 0.051 & 2 & 2 & 2 & 4 & 0.051 & 3 & 2 & 2 & 4 & 0.061 \\
\hline
1 & 2 & 3 & 1 & 0.036 & 2 & 2 & 3 & 1 & 0.109 & 3 & 2 & 3 & 1 & 0.029 \\
\hline
1 & 2 & 3 & 2 & 0.051 & 2 & 2 & 3 & 2 & 0.054 & 3 & 2 & 3 & 2 & 0.032 \\
\hline
1 & 2 & 3 & 3 & 0.067 & 2 & 2 & 3 & 3 & 0.017 & 3 & 2 & 3 & 3 & 0.036 \\
\hline
1 & 2 & 3 & 4 & 0.051 & 2 & 2 & 3 & 4 & 0.054 & 3 & 2 & 3 & 4 & 0.032 \\
\hline
1 & 2 & 4 & 1 & 0.102 & 2 & 2 & 4 & 1 & 0.055 & 3 & 2 & 4 & 1 & 0.069 \\
\hline
1 & 2 & 4 & 2 & 0.052 & 2 & 2 & 4 & 2 & 0.051 & 3 & 2 & 4 & 2 & 0.061 \\
\hline
1 & 2 & 4 & 3 & 0.017 & 2 & 2 & 4 & 3 & 0.047 & 3 & 2 & 4 & 3 & 0.055 \\
\hline
1 & 2 & 4 & 4 & 0.052 & 2 & 2 & 4 & 4 & 0.051 & 3 & 2 & 4 & 4 & 0.061 \\
\hline
1 & 3 & 1 & 1 & 0.048 & 2 & 3 & 1 & 1 & 0.041 & 3 & 3 & 1 & 1 & 0.253 \\
\hline
1 & 3 & 1 & 2 & 0.049 & 2 & 3 & 1 & 2 & 0.043 & 3 & 3 & 1 & 2 & 0.118 \\
\hline
1 & 3 & 1 & 3 & 0.048 & 2 & 3 & 1 & 3 & 0.041 & 3 & 3 & 1 & 3 & 0.008 \\
\hline
1 & 3 & 1 & 4 & 0.049 & 2 & 3 & 1 & 4 & 0.043 & 3 & 3 & 1 & 4 & 0.118 \\
\hline
1 & 3 & 2 & 1 & 0.048 & 2 & 3 & 2 & 1 & 0.050 & 3 & 3 & 2 & 1 & 0.060 \\
\hline
1 & 3 & 2 & 2 & 0.049 & 2 & 3 & 2 & 2 & 0.051 & 3 & 3 & 2 & 2 & 0.062 \\
\hline
1 & 3 & 2 & 3 & 0.048 & 2 & 3 & 2 & 3 & 0.050 & 3 & 3 & 2 & 3 & 0.060 \\
\hline
1 & 3 & 2 & 4 & 0.049 & 2 & 3 & 2 & 4 & 0.051 & 3 & 3 & 2 & 4 & 0.062 \\
\hline
1 & 3 & 3 & 1 & 0.048 & 2 & 3 & 3 & 1 & 0.052 & 3 & 3 & 3 & 1 & 0.008 \\
\hline
1 & 3 & 3 & 2 & 0.049 & 2 & 3 & 3 & 2 & 0.041 & 3 & 3 & 3 & 2 & 0.118 \\
\hline
1 & 3 & 3 & 3 & 0.048 & 2 & 3 & 3 & 3 & 0.052 & 3 & 3 & 3 & 3 & 0.253 \\
\hline
1 & 3 & 3 & 4 & 0.049 & 2 & 3 & 3 & 4 & 0.041 & 3 & 3 & 3 & 4 & 0.118 \\
\hline
1 & 3 & 4 & 1 & 0.049 & 2 & 3 & 4 & 1 & 0.050 & 3 & 3 & 4 & 1 & 0.060 \\
\hline
1 & 3 & 4 & 2 & 0.039 & 2 & 3 & 4 & 2 & 0.051 & 3 & 3 & 4 & 2 & 0.062 \\
\hline
1 & 3 & 4 & 3 & 0.049 & 2 & 3 & 4 & 3 & 0.050 & 3 & 3 & 4 & 3 & 0.060 \\
\hline
1 & 3 & 4 & 4 & 0.039 & 2 & 3 & 4 & 4 & 0.051 & 3 & 3 & 4 & 4 & 0.062 \\
\hline

\end{tabular}
\caption{The LOCC protocol for target state $\rhot{UNF3}^{1.0}$.}
\label{table:rhounf5}

\end{table*}

\end{appendix}

\clearpage

\bibliography{bibliography}

\end{document}